\documentclass[12pt]{article}
\usepackage[T1]{fontenc}
\usepackage[utf8]{inputenc}
\usepackage{lmodern}
\usepackage{jheppub}

\usepackage{amssymb,amsfonts}
\usepackage{parskip}

\usepackage{natbib}
\setcitestyle{square, comma, numbers, compress}
\usepackage{wrapfig}
\usepackage{epsfig}
\usepackage{float}
\usepackage{placeins}
\usepackage{dsfont}
\usepackage{arydshln}
\usepackage{caption}
\usepackage{todonotes}
\usepackage{extarrows}
\usepackage{tensor}
\usepackage{slashed}
\usepackage{enumitem}
\usepackage{mathtools}

\usepackage{xcolor}
\usepackage{tikz}
\usetikzlibrary{arrows.meta,positioning,calc,decorations.pathmorphing}
\usetikzlibrary{3d}
\usetikzlibrary{calc, decorations.markings}

\pgfkeys{/tikz/.cd, view angle/.initial=0, view angle/.store in=\picangle}

\tikzset{
    horizontal/.style={y={(0,sin(\picangle))}},
    vertical at/.style={x={([horizontal] #1:1)}, y={(0,cos(\picangle)cm)}},
    every label/.style={font=\tiny, inner sep=1pt},
    shorten/.style={shorten <=#1, shorten >=#1},
    shorten/.default=3pt,
    ->-/.style={decoration={markings, mark=at position #1 with {\arrow{>}}}, postaction={decorate}},
    ->-/.default=0.5,
    dark plane/.style={rotate around x=\angledp,canvas is xy plane at z=0,draw=cyan,fill=cyan!25},
}

\newcommand{\nn}{\nonumber}

\newcommand{\ket}[1]{|#1\rangle}
\newcommand{\bra}[1]{\langle#1|}
\newcommand{\qc}{\,,\qquad}

\newcommand{\Z}{{\mathds Z}}

\def\implies{\Rightarrow}

\def\calG{\mathcal{G}}
\def\calS{\mathcal{S}}
\def\calT{\mathcal{T}}

\newcommand{\cH}{\mathcal{H}}

\newcommand{\cG}{\mathcal{G}}
\newcommand{\cV}{\mathcal{V}}
\newcommand{\cA}{\mathcal{A}}
\newcommand{\cR}{\mathcal{R}}
\newcommand{\cT}{\mathcal{T}}
\newcommand{\cS}{\mathcal{S}}
\newcommand{\cN}{\mathcal{N}}
\newcommand{\cM}{\mathcal{M}}
\newcommand{\cL}{\mathcal{L}}

\def\lae{\mathrel{\mathop{\smash{\lower .5 ex \hbox{$\stackrel<\sim$}}}}}
\def\lae{\mathrel{\mathop{\smash{\lower .5 ex \hbox{$\stackrel>\sim$}}}}}

\title{Non-Invertible Symmetries and Boundaries for Two-Dimensional Fermions}

\author{Guillermo Arias-Tamargo$^a$,}
\author{Philip Boyle Smith$^{b,c}$,}
\author{Rishi Mouland$^a$,}
\author{and Maxwell L. Velásquez Cotini Hutt$^{a,d}$}

\affiliation[a]{Abdus Salam Centre for Theoretical Physics, \\Imperial College London, London SW7 2AZ, UK}
\affiliation[b]{SISSA, via Bonomea 265, 34136 Trieste, Italy}
\affiliation[c]{INFN, Sezione di Trieste, via Valerio 2, 34127 Trieste, Italy}
\affiliation[d]{Department of Applied Mathematics and Theoretical Physics, \\University of Cambridge, CB3 0WA, UK\\}

\emailAdd{g.arias-tamargo@imperial.ac.uk}
\emailAdd{philip.boyle.smith@sissa.it}
\emailAdd{r.mouland@imperial.ac.uk}
\emailAdd{mv623@cam.ac.uk}

\abstract{
We study the relation between boundary conditions and categorical symmetries of two-dimensional fermionic conformal field theories. We determine all anomaly-free invertible global symmetries of two free complex Weyl fermions, which take the form $\mathbb{Z}_k$ for each primitive Pythagorean triple $a^2 + b^2 = k^2$. The theory is self-dual under gauging any of these symmetries, and so to each there is associated a non-invertible topological defect. We study the properties of these lines, and show that any conformal boundary condition of two Dirac fermions that preserves a $U(1)^2$ symmetry can be found by dressing a trivial Dirichlet boundary with one of them. We discuss two microscopic descriptions of these defects: fermions coupled to a quantum-mechanical rotor degree of freedom; and an abelian gauge theory that realises symmetric mass generation in a half-space.
}

\begin{document}
\pagestyle{plain} \setcounter{page}{1}
\newcounter{bean}
\baselineskip16pt \setcounter{section}{0}

\maketitle

\section{Introduction}

Interesting features of quantum field theories (QFTs) can emerge when the theory is placed on a manifold with boundary. In understanding the fate of bulk excitations incident on such boundaries, and their interaction with boundary degrees of freedom that may live there, global symmetries provide a vital tool. In particular, the 't Hooft anomalies of these symmetries provide robust insight beyond the reaches of perturbative methods. As a first port of call, the presence of such an anomaly is often effective in ruling out scenarios. More subtle is the question of what we can learn from the \emph{absence} of an anomaly.

Consider for instance a Conformal Field Theory (CFT) with symmetry $G$ on a manifold with boundary. When does there exist a conformal boundary condition preserving $G$? One is in particular interested in \emph{simple} conformal boundaries, which for scattering means that the boundary cannot absorb $G$ charge. It has been appreciated that such a boundary condition does not exist whenever $G$ has a 't Hooft anomaly \cite{Jensen_2018, Thorngren_2021}. One is led to consider whether the converse could be true: that if $G$ is anomaly-free, then a $G$-symmetric simple conformal boundary \emph{does} exist. While a number of non-trivial such boundary conditions have been constructed (see e.g.\ \cite{Han:2017hdv,Li:2022drc,BoyleSmith:2019jnh,BoyleSmith:2020nuf,Numasawa:2017crf,Thorngren_2021}), a few counterexamples in two dimensions have recently emerged \cite{Choi:2023xjw,Wei:2025zyd}.

The non-triviality of this question is most simply seen in the theory of two Dirac fermions in two spacetime dimensions. Consider the $G=U(1)$ symmetry under which left-movers carry charges $3$ and $4$, while right-movers carry charges $5$ and $0$. This symmetry has no anomaly, by virtue of
\begin{align}
3^2 + 4^2 = 5^2 + 0^2 \,.
\end{align}
One can, in this case, define the theory on the half-line $x>0$ with a simple conformal boundary at $x=0$ that preserves this $U(1)$. Given such a boundary condition, if we send in a left-mover of charge $3$, what comes back? Charge conservation requires a state of charge $3$, but there are no right-moving local operators of that charge.
\begin{center}
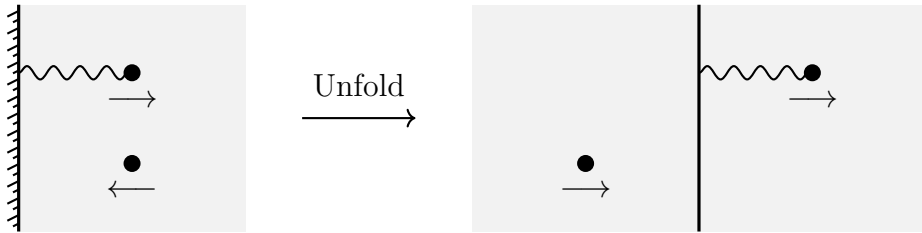
\vspace{1em}
\begin{minipage}{0.8\textwidth}\centering 
    \begin{tikzpicture}[scale=0.75]
        \fill[black!5] (6,0) rectangle (10,-4);
        \draw[very thick] (6,0) -- (6,-4);
        \fill (8,-1.2) circle (0.15);
        \fill (8,-2.8) circle (0.15);
        \draw[decorate, decoration=snake,thick] (6,-1.2) -- (8,-1.2);
        \node[below] at (8,-1.4) {$\longrightarrow$};
        \node[below] at (8,-3) {$\longleftarrow$};
        \foreach \x in {0,...,12} \draw[thick] (6-.2,-4+.3*\x+.04+.15) -- (.2+6-.2,-4+.3*\x+.1+.04+.15);
        \foreach \x in {-1,...,11} \draw[thick] (6-.1,-4+.3*\x+.09+.15+.15) -- (.2+6-.2,-4+.3*\x+.1+.04+.15+.15);
        \draw[thick,->] (11,-2) -- (13,-2);
        \node[above] at (12,-1.8) { Unfold};
        \begin{scope}[shift={(12,0)}]
            \fill[black!5] (2,0) rectangle (6,-4);
            \fill[black!5] (6,0) rectangle (10,-4);
            \draw[very thick] (6,0) -- (6,-4);
             \fill (8,-1.2) circle (0.15);
            \fill (4,-2.8) circle (0.15);
            \draw[decorate, decoration=snake,thick] (6,-1.2) -- (8,-1.2);
            \node[below] at (8,-1.4) {$\longrightarrow$};
            \node[below] at (4,-3) {$\longrightarrow$};
        \end{scope}	
    \end{tikzpicture} 
    \captionof{figure}{Scattering off a conformal boundary for Dirac fermions that preserves some chiral symmetry, and its unfolded version. }\label{fig: intro}
\end{minipage}\vspace{1em}
\end{center}
The known resolution of this apparent puzzle is that a state created by a local, left-moving fermion operator is generically scattered by the boundary into a right-moving state created not by a local operator, but rather by a twist operator that is attached to the boundary by a topological line. This is shown in the left-hand panel of Figure~\ref{fig: intro}. The lines that are generated belong to a particular anomaly-free $\Z_5$ symmetry. These lines are precisely such that their endpoints effectively harbour right-moving free fermions with charges $3$ and $4$. The smoking gun of such an outgoing twisted sector state is that, while we preserve the anomaly-free $U(1)$ symmetry we wanted to preserve, this state carries fractional charge under some other anomalous symmetry.

These boundary conditions (and their generalisations to $N>2$ Dirac fermions) are relevant in many contexts across physics, from the Kondo problem in condensed matter theory \cite{affleck1995} to the study of D-branes in superstring theory \cite{Recknagel_1999, Gaberdiel:2002my, gaberdiel2003boundary}. They are also crucial in understanding what happens when we scatter massless fermions off heavy monopoles in four-dimensional gauge theory, which has been argued to be captured at low energies by a $(1+1)$-dimensional boundary CFT problem of chiral fermions describing the lowest angular momentum modes \cite{Callan:1982ah,Rubakov:1982fp,Callan:1982ac,Callan:1982au,Affleck:1993np,Yegulalp:1994eq,Maldacena_1997,vanBeest:2023dbu,vanBeest:2023mbs}.

It is useful to consider the fate of the boundary condition shown in Figure~\ref{fig: intro} when we `unfold' along $x=0$, as shown in the right-hand panel. We find two Weyl fermions on the infinite line, with a timelike defect at $x=0$. This defect has the property that local operators become attached to a $\Z_5$ topological line when they pass through it, the other end of which remains on the defect. This is precisely the hallmark of a non-invertible $0$-form symmetry defect placed at $x=0$.

The purpose of this paper is to make manifest this relationship between chiral boundary conditions and non-invertible topological defects.

We will show that the theory of two Weyl fermions admits a non-invertible topological line defect associated with each primitive Pythagorean triple
\begin{align}
a^2 + b^2 = k^2 \,.
\label{eq: triple}
\end{align}
To do so, we identify a $\Z_k$ symmetry of the theory for each such triple, and establish a duality between the theory with and without this symmetry gauged. Performing this gauging in a half-space then generates the corresponding non-invertible line defect.

Our construction is exhaustive: we find all self-duality defects of the theory that arise from gauging an invertible symmetry, up to fusion with invertible lines. We study the fusion properties of these self-duality defects. In particular we prove that, by fusing with suitable invertible lines, we can find a self-duality defect for each $(a,b,k)$ which belongs to the finite fusion category first considered by Tambara and Yamagami \cite{Tambara:1998vmj}. 


We then show that a conformal boundary condition of two Dirac fermions preserving some generic anomaly-free $U(1)^2$ symmetry\footnote{We focus on preserving a $U(1)^2$, as opposed to a single $U(1)$, as then the scattering is uniquely solved. For the 3450 model this amounts to also imposing the preservation of a second $U(1)$ symmetry which one can take to have charges $(4,-3,0,-5)$ or $(4,-3,0,5)$. Only in the former case is the conformal boundary condition simple, while in the latter it harbours an unpaired Majorana mode.} can be constructed from a suitable non-invertible line in the theory of two Weyl fermions, which we identify. For instance, for the 3450 model, unsurprisingly we need a $(a,b,k)=(3,4,5)$ line. We provide two essentially equivalent constructions of this boundary condition: folding two Weyl fermions along a non-invertible line; and dressing a Dirichlet boundary preserving a vector-like symmetry of two Dirac fermions with such a line. This boundary condition is only simple for half of the choices of $U(1)^2$, while for the other half it harbours a single, unpaired Majorana mode. 

Finally, we exhibit two UV descriptions of the non-invertible topological lines. The first of these consists of bulk Weyl fermions coupled to a rotor impurity localised at $x=0$. In the context of fermion-monopole scattering, this description can be traced back to the rotor degree of freedom of a 't Hooft--Polyakov monopole, as first considered by Polchinski \cite{Polchinski:1984uw}.

Secondly, we provide a UV setup that flows to each of the $U(1)^2$-preserving boundary conditions discussed above. This consists of a gauge theory that enacts \textit{symmetric mass generation} (SMG), whereby two free Dirac fermions are deformed to a trivially gapped phase while preserving a chiral but necessarily anomaly-free $U(1)^2$ symmetry. By doing so in a spatially-modulated way, we are able to dynamically generate a conformal boundary condition preserving any such $U(1)^2$. In particular this construction sheds light on the origin of the Majorana mode required for half of the boundary conditions.
 
The rest of the paper is organised as follows. In section~\ref{sec:anomalies} we classify all anomaly-free invertible symmetries of the theory of two Weyl fermions and demonstrate self-duality under gauging all such symmetries. In section~\ref{sec: Non-invertible interfaces and boundaries} we construct a family of non-invertible self-duality defects associated to each of these symmetries. We study the fusion of these lines, as well as the scattering of local excitations through them. In section~\ref{sec: Symmetric boundary conditions} we use these non-invertible lines to construct conformal boundaries for two Dirac fermions that preserve a generic anomaly-free $U(1)^2$ symmetry. We then discuss some microscopic setups which flow to these non-invertible defects at low energies in section~\ref{sec: UV}. Finally, we provide some discussion in section~\ref{sec: discussion}.

\textbf{Note:} This paper was submitted in coordination with \cite{christian}, the results of which have a partial overlap with those of this work.

\section{Discrete gauging and self-dualities}
\label{sec:anomalies}

We begin simply, with the theory of two complex Weyl fermions in two dimensions,
\begin{align}
  \mathcal{L} = i \psi_1^\dagger \partial_+ \psi_1 + i \psi_2^\dagger \partial_+ \psi_2\,,
  \label{eq: Weyl fermions}
\end{align}
where we take our two single-component Weyl spinors $\psi_1,\psi_2$ to be right-moving, and so $\partial_+ = \partial_t + \partial_x$. We call this theory $\cT$.

The aim of this section will be to gauge a symmetry $\cG$ of $\cT$. Such a symmetry must be anomaly-free; our first result will be a classification of all such symmetries. They are all finite symmetries, and indeed are all cyclic groups of odd order.

We will then gauge a generic such anomaly-free cyclic $\cG$, which raises some subtleties. Coupling background fields for $\cG$ requires a choice of $\cG$-action not just on the local fields $\psi_{1,2}$, corresponding to the states of the untwisted Neveu--Schwarz (NS) sector, but also on the $\cG$-twisted NS and Ramond (R) sectors, as well as a choice of $(-1)^F$ assignment on these new states. These choices reflect the freedom to add counter-terms for the background $\cG$ gauge field. We will see explicitly that most such couplings to $\cG$ will be inconsistent, in the sense that attempting to gauge $\cG$ will not give rise to a well-defined fermionic CFT. However, given that $\cG$ is anomaly-free, there must be at least one consistent coupling. On general grounds, different choices of consistent couplings are expected to differ by stacking SPT phases for the $\cG$ symmetry, which are classified by the reduced cobordism group $(\widetilde{I_\Z \Omega^\text{Spin}})^3(B\cG)$. For an odd cyclic group $\cG$, this group is trivial. Thus there should be a unique way of gauging $\cG$. We determine this choice.

Finally we show that the resulting gauge theory $\cT/\cG$ is precisely the theory we started with: two free Weyl fermions. This had to happen, since gauging a discrete symmetry does not change the central charges, and $\cT$ is the unique fermionic CFT with central charges $(c, \bar{c}) = (0, 2)$.\footnote{More precisely, it is the unique \emph{single-vacuum} such theory. Since our symmetry $\cG$ is faithful, gauging $\cG$ preserves the fact the theory has a single vacuum, i.e.\ does not add a decoupled TQFT. Note we will also later encounter the theory $\cT\otimes \text{Arf}$, but as we shall see, this is isomorphic to $\cT$.} However, in demonstrating this self-duality under gauging $\cG$ explicitly, we will establish how both local and extended operators are mapped by the gauging, along with the global symmetries that act on them. This mapping will be key to understanding scattering and boundary conditions later.

Before we look at Weyl fermions, however, we provide a brief interlude on an even simpler model---a single compact boson---where many of these same subtleties arise.

\subsection{An interlude: the compact boson}

As a warm-up, consider a compact boson in two dimensions at radius $R$,
\begin{align}
  \cL = \frac{R^2}{4\pi}(\partial \phi)^2,\qquad \phi \sim \phi+2\pi \,.
  \label{eq: compact boson}
\end{align}
This theory has a $U(1)_\text{shift}\times U(1)_\text{winding}$ global symmetry. Suppose we want to gauge $\cG=\Z_2\subset U(1)_\text{shift}$. This $\cG$ has a mixed anomaly with $U(1)_\text{winding}$. This can be seen by writing the partition function $Z_\alpha$ of the theory on $S^1$ with boundary conditions $e^{i\phi(\sigma + 2\pi)} =(-1)^\alpha e^{i\phi(\sigma)} $, where $\sigma$ is the $S^1$ coordinate and $\alpha=0,1$ correspond to the untwisted sector and the $\Z_2$-twisted sector, respectively. Dressing this partition function with fugacities for $U(1)_\text{shift}$ and $U(1)_\text{winding}$, we have
\begin{align}
  Z_\alpha(R;\tau,z_\text{s},z_\text{w}) &= \text{Tr}_{\cH_\alpha} \, \zeta^{h-1/24} \bar{\zeta}^{\bar{h}-1/24} z_\text{s}^{Q_\text{s}} z_\text{w}^{Q_\text{w}} \nn\\
  &=\frac{1}{|\eta(\tau)|^2} \sum_{\substack{ n\in\Z \\ m \in\Z-\frac{\alpha}{2}}} \zeta^{\frac{1}{2}\left(\frac{n}{R}-m R\right)^2}\bar\zeta^{\frac{1}{2}\left(\frac{n}{R}+m R\right)^2}z_\text{s}^{n}z_\text{w}^{m} \,.
\end{align}
Here, $\zeta=e^{2\pi i \tau}$ where $\tau$ is the modular parameter of the torus, and $\eta(\tau)$ is the Dedekind eta function. Meanwhile, $Q_\text{s}$ and $Q_\text{w}$ are the generators of $U(1)_\text{shift}$ and $U(1)_\text{winding}$, respectively.

The mixed anomaly between $\Z_2\subset U(1)_\text{shift}$ and $U(1)_\text{winding}$ is then diagnosed by the projective representation of $U(1)_\text{winding}$ carried by the $\Z_2$-twisted sector: under $z_w\to e^{2\pi i}z_w$, the partition function picks up a sign. Only a double cover acts non-projectively.
As a result, there is a two-fold ambiguity in how we fix the action of $\cG$: should we take it to be generated by $e^{i\pi Q_\text{s}}$, or by $e^{i\pi Q_\text{s}+2\pi i Q_\text{w}}$? These two choices coincide in the untwisted sector, but differ by a sign in the twisted sector.

Let us parameterise these two choices by $t\in \{0,1\}$. Turning off fugacities for brevity, the resulting gauge theory partition function is
\begin{align}
  Z^\text{gauged}_t(R;\tau) &= \frac{1}{2}\sum_{\alpha,\beta\in \{0,1\}} Z_\alpha (R;\tau,e^{\pi i \beta},e^{2\pi i t \beta })		\nn\\
  &= \frac{1}{2}\sum_{\alpha,\beta\in \{0,1\}} (-1)^{t\alpha\beta} Z_\alpha (R;\tau,e^{\pi i \beta},1)			\nn\\
  &= \left\{ 
  \begin{aligned}
  	\,\,& Z_{0}(R/2;\tau,1,1)\, ,	\hspace{80mm}&t=0		\\
  	& \frac{1}{|\eta(\tau)|^2} \sum_{n,m\in \Z} \zeta^{\left((n+m)R+2(n-m)/R\right)^2/16} \bar{\zeta}^{\left((n+m)R-2(n-m)/R\right)^2/16}\, , & t=1
  \end{aligned}
  \right.
\end{align}
where the sum over $\beta$ enacts the projection onto $\cG$ singlets. 

The case $t=0$ recovers the familiar result that gauging $\Z_2\subset U(1)_\text{shift}$ has the intuitive effect of halving the radius of the compact boson. In particular, this defines a self-duality of the theory at $R^2=2$ under gauging $\cG$ by virtue of T-duality $R\leftrightarrow 1/R$. Self-dualities of the compact boson CFT and the associated categorical symmetries have been discussed in \cite{Fuchs:2007tx,Bachas:2012bj,Bachas:2013ora,Thorngren:2021yso,ChoiCordovaHsinLam2021,Damia2024,Argurio:2024ewp,Niro:2022ctq,Bharadwaj:2024gpj}.

The $t=1$ case is more interesting. Although we started with a bosonic theory, the gauged theory has spins $(h-\bar{h}) = (n^2 - m^2)/2\in \frac{1}{2}\Z$, and is thus not a well-defined bosonic CFT. 
Indeed, we recognise $Z^\text{gauged}_{t=1}(R;\tau)$ as nothing but the NS-NS partition function of the radius $R$ theory on the `Dirac branch' of the $c=1$ fermionic moduli space \cite{Karch:2019lnn}. The free Dirac fermion is recovered at $R^2=2$. We see then that it only makes sense to gauge $\cG=\Z_2$ with $t=1$ if we also specify a spin structure. In doing so, we are performing fermionisation.

From a Lagrangian perspective, the additional phase $(-1)^{t\alpha\beta}$ corresponds to our ability to add an additional counter-term when gauging the discrete symmetry $\cG$. In this interpretation, one identifies the phase $(-1)^{\alpha\beta}=(-1)^{\text{Arf}[\rho+a]}$ where $a\in H^1(M,\cG)$ is the $\cG\cong\Z_2$ gauge field and $\rho$ is the NS-NS spin structure, which couple via the Arf invariant \cite{Karch:2019lnn}.

In some contexts one might refer to the phase $(-1)^{t\alpha\beta}$ as a choice of \textit{discrete torsion}. The different choices of discrete torsion compatible with modular invariance for bosonic theories are labelled by elements of $(\widetilde{I_\Z \Omega^\mathrm{SO}})^3(B\cG) \cong H^2(\cG; U(1))$ \cite{Vafa:1986wx}, and here $H^2(\Z_2; U(1))=0$. Indeed, we see that the gauged theory partition function is only modular invariant for the unique choice $t=0$. But the choices of discrete torsion compatible with modular invariance for fermionic theories are classified by $(\widetilde{I_\Z \Omega^\mathrm{Spin}})^3(B\cG) \cong H^2(\cG; U(1)) \oplus H^1(\cG; \Z_2) \cong \Z_2$, for $\cG\cong\Z_2$. Thus when we enlarge to the class of fermionic theories, both choices $t = 0, 1$ become allowed.

Below, we will see how an analogous phenomenon occurs in the theory of 2 Weyl fermions, whereby the presence of a mixed 't Hooft anomaly allows us to introduce an additional phase when gauging a discrete group. Similarly to the compact boson CFT, we will find that there is a unique choice of phase which will lead to a well-defined fermionic theory after gauging, whereas other choices of phases would correspond to enlarging to the class of parafermionic theories \cite{Fateev:1985mm}. Determining this phase will be essential to find a self-duality. In this case, unlike for the compact boson, the naïve choice won't be the one that does the job.

\subsection{Anomaly-free symmetries of Weyl fermions}\label{sec: classifying groups}

Let us return to the theory $\cT$ of two Weyl fermions \eqref{eq: Weyl fermions}. Our goal is to gauge a discrete invertible 0-form symmetry $\cG$. So let us first classify all such anomaly-free symmetries.

The set of all invertible 0-form symmetries of $\cT$ forms the group $SO(4)$.\footnote{Here we use the convention that the symmetry of a fermionic theory is the quotient that acts faithfully on the NS sector. The proof that there are no further symmetries than $SO(4)$ is as follows. Any invertible symmetry acts on the spin-1 currents, which form the $\mathfrak{so}(4)$ Lie algebra under fusion, as an automorphism. Such automorphisms are either inner (conjugation by an element of $SO(4)$), or outer (conjugation by an element of $O(4) \backslash SO(4)$). Since the theory is rational with respect to the $\widehat{\mathfrak{so}(4)}_1$ currents, the possible extensions of these automorphisms to full symmetries of the theory are tightly constrained. The inner automorphisms extend to $SO(4)$ symmetries, while the outer automorphisms extend to non-symmetry interfaces $\cT \to \cT \otimes \text{Arf}$ labelled by elements of $O(4) \backslash SO(4)$. These will be discussed further in section~\ref{subsec: self-duality}.}  The starting point of the classification is the isomorphism
\begin{align}
  SO(4) \cong (SU(2)_L \times SU(2)_R) / \Z_2\,,
\end{align}
where an element $(\theta_L, \theta_R)$ of the Cartan of $SU(2)_L \times SU(2)_R$ acts on the fermions as
\begin{align*}
  \psi_1 \to e^{i(\theta_L + \theta_R)} \psi_1  \,,\\
  \psi_2 \to e^{i(\theta_L - \theta_R)} \psi_2 \,,
\end{align*}
which explains why $(\theta_L, \theta_R) = (\pi, \pi)$ acts trivially and is quotiented out.

We first claim that any element of either $SU(2)_L$ or $SU(2)_R$ of finite order $n > 1$ generates an anomalous $\Z_n$ symmetry. To prove this, without loss of generality we can rotate the element into the Cartan of the $SU(2)$ and raise it to a suitable power so that it acts as
\begin{align}
  \psi_1 &\rightarrow e^{2\pi i/n} \psi_1 \,, \\
  \psi_2 &\rightarrow e^{\pm 2\pi i/n} \psi_2 \,.
\end{align}
Then the condition for the vanishing of the anomaly for such a $\mathbb{Z}_n$ symmetry depends on the parity of $n$. For $n$ odd, the anomaly is given by a straightforward mod $n$ reduction of any $U(1)$ in which we embed it. In contrast, for $n$ even there is an additional non-perturbative anomaly, such that the anomaly is actually mod $2n$ valued \cite{vanBeest:2023dbu}. To summarise, the anomaly-free condition is
\begin{align}
  1^2 + (\pm 1)^2 = 0 \text{ mod }
  \begin{cases}
    n  & n \text{ odd} \\
    2n  & n \text{ even}
  \end{cases}\,,
\end{align}
which is false for all $n > 1$. Hence any such $\Z_n$ is anomalous.

Now suppose $\cG \subseteq SO(4)$ is a finite anomaly-free symmetry. The above result forces the projection maps of $\cG$ onto the left and right $SU(2)/\Z_2 \cong SO(3)$ factors to be injective.\footnote{Suppose $(g_L, g_R) \in \cG \subseteq (SU(2)_L \times SU(2)_R) / \Z_2$ is in the kernel of the projection $\cG \to SU(2)_L/\Z_2$. Then $g_L = \pm 1$. Using the $\Z_2$ quotient, without loss of generality $g_L = 1$. Thus $\cG$ contains $(1, g_R) \in SU(2)_R$, which generates an anomalous subgroup unless $g_R = 1$.} The existence of these injective maps then means $\cG$ can be regarded as a finite subgroup of $SO(3)$. Such subgroups have an ADE classification:
\begin{itemize}

\item The cyclic groups $\cG \cong \Z_n$

Any faithfully-acting $\Z_n$ symmetry can be taken to be generated by
\begin{align}
    \psi_1 &\to e^{ 2\pi i q/n} \psi_1  \,,\\
    \psi_2 &\to e^{-2\pi i p/n} \psi_2 \,,
\label{eq: Zk action}
\end{align}
for some integers $p, q$ with $\gcd(p, q) = 1$.\footnote{A generic generator of $\Z_n$ will be of this form with $\gcd(p,q,n)=1$, but one can show that a generator can always be chosen with the stronger property $\gcd(p,q)=1$.} To then study the anomaly of this $\Z_n$ it is useful to define
\begin{align}
  k \equiv p^2 + q^2 = \left\{ \begin{aligned}
    1 \mod 4 \qquad &\text{if }pq\text{ even} \\
    2 \mod 4 \qquad &\text{if }pq\text{ odd}
  \end{aligned} \right.\,\,.
\label{eq: k mod 4}
\end{align}
Note in particular that $k$ is coprime to both $p$ and $q$. As discussed above, if $n$ is odd, the anomaly is given by a straightforward mod $n$ reduction of any $U(1)$ in which we embed it.
Thus for $n$ odd, $\Z_n$ is anomaly-free precisely if we have
\begin{align}
  k=0 \mod n \,.
\end{align}
For each coprime pair $(p,q)$ there exists a maximal odd $n$ satisfying this equation. All other solutions for $n$ are precisely the divisors of this maximal $n$, and thus define subgroups of a larger anomaly-free group; we will say more about these subgroups below. The maximal $n$ is given by
\begin{align}
  n = \frac{k}{\gcd(k,2)} = \left\{ \begin{aligned}
    &k \qquad  &&\text{if }pq\text{ even} \\
    &\frac{k}{2}  \qquad &&\text{if }pq\text{ odd}
  \end{aligned} \right.\,\,.
\label{eq: n}
\end{align}
Conversely for $n$ even, $\Z_n$ is only free of anomalies if the following stronger constraint is satisfied \cite{vanBeest:2023dbu}
\begin{align}
  k=0 \mod 2n \,.
\end{align}
One sees from \eqref{eq: k mod 4} that no choice of $(p,q,n)$ with even $n$ satisfies this equation. Thus, there are no anomaly-free $\Z_n$ symmetries with $n$ even.

\item The non-abelian groups $\cG \cong \mathbb{D}_{2n}, \mathbb{A}_4, \mathbb{S}_4, \mathbb{A}_5$

All these groups have a $\Z_2$ subgroup. By the cyclic case, all $\Z_2$ symmetries are anomalous. Therefore these symmetries are anomalous too.

\end{itemize}
In summary, we find an anomaly-free $\Z_n$ for each coprime pair $p,q\in \Z$ with $n$ given in \eqref{eq: n}. We denote this symmetry $\cG_{(p,q)}$. Symmetries of this type then fall into two types:
\begin{itemize}
  \item For each coprime pair $(p,q)$ with $pq$ even, we find anomaly-free $\cG_{(p,q)}\cong \Z_k$ with $k=p^2 + q^2$, generated by
\begin{align}
    \psi_1 \to e^{2\pi i q/k}\psi_1\,,\quad \psi_2\to e^{-2\pi i p/k}\psi_2\,.
\label{eq: A action}
\end{align}
\item For each coprime pair $(p,q)$ with $pq$ odd, we find anomaly-free $\cG_{(p,q)}\cong \Z_{k/2}$ with $k = p^2 + q^2$, generated by
\begin{align}
    \psi_1 \to e^{4\pi i q/k}\psi_1\,,\quad \psi_2\to e^{-4\pi i p/k}\psi_2\,.
\end{align}
\end{itemize}
There are some small subtleties to address. Firstly note that $\cG_{(p,q)}$ is trivial when $p = 0$, or $q = 0$, or $|p| = |q| = 1$, so we can exclude these cases. Next note that there is an eight-fold degeneracy in  the parameterisation of the symmetry groups $\cG_{(p,q)}$ in terms of coprime pairs $p,q\in \Z$. Firstly within each of the two types the following groups are equivalent,
\begin{align}
\cG_{(p,q)}\cong \cG_{(-p,-q)}\cong \cG_{(q,-p)} \cong \cG_{(-q,p)}\,,\qquad \text{for all }p,q\text{ coprime} \,.
\label{eq: group equivalences}
\end{align}
Secondly, every group of the first type is equivalent to one of the second, and vice versa. These equivalences are
\begin{align}
\cG_{(p,q)}\cong \cG_{(p-q,p+q)}\,,\qquad \text{for all }p,q\text{ coprime with }pq\text{ even} \,.
\label{eq: A B equivalence}
\end{align}
These exhaust all equivalences between the maximal anomaly-free groups, as is proven in Appendix~\ref{app: proofs}. One can then take as the distinct maximal anomaly-free invertible symmetries the set $\{\cG_{(p,q)}\}$ with $pq$ even and $p,q>0$.

Finally, we need to consider subgroups of the $\cG_{(p,q)}$, where it suffices to consider $pq$ even as we've just seen. These subgroups are automatically anomaly-free. It turns out that any such subgroup is already included in the classification. Precisely, if $k$ has a prime factor $d$, then the subgroup $\Z_{k/d}\subset \cG_{(p,q)}\cong \Z_k$ is equivalent to $\cG_{(p',q')}$ for some $p',q'$ coprime with $p'q'$ even. A proof of this fact is also found in Appendix~\ref{app: proofs}.

We conclude that the set of \emph{all} inequivalent anomaly-free invertible symmetries of two Weyl fermions is given precisely by
\begin{align}
\{\cG_{(p,q)}\cong \Z_k\}_{(p,q)\in \cS}\,,
\end{align}
where, as ever, $k=p^2 + q^2$, $\cG_{(p,q)}$ is generated by
\begin{align}
    \psi_1 \to e^{2\pi i q/k}\psi_1\,,\quad \psi_2\to e^{-2\pi i p/k}\psi_2\,,
\end{align}
and
\begin{align}
\cS = \{(p,q)\in \Z_{>0}^2: \gcd(p,q)=1,\,\, pq \text{ even} \} \,.
\label{eq: S}
\end{align}
Note that there is an equivalent parameterisation of this set which will prove useful later, in terms of primitive Pythagorean triples:
\begin{align}
\begin{split}
    \cS = \{(a,b,k) \in \Z_{\neq 0}^2 \times \Z_{>0} :{} &a^2 + b^2 = k^2,\,\, \gcd(a,b,k)=1,\\
    &(a,b,k)\sim (b,a,k)\sim (-a,-b,k) \} \,.
\end{split}
\label{eq: S Pythagorean}
\end{align}
The bijection between elements of the presentation \eqref{eq: S} and representatives from each equivalence class in the presentation \eqref{eq: S Pythagorean} was given by Euclid,
\begin{align}
    a = p^2 - q^2,\quad b = 2pq,\quad k =p^2 + q^2 \,.
\label{eq: Euclid}
\end{align}

\subsection{Gauging cyclic symmetries}\label{sec:mixed_anomalies_U1s}

Let us now fix some $\cG=\cG_{(p,q)}$ with $p,q$ coprime and $pq$ even, and consider gauging this symmetry.

Following our example of the compact boson, the first step is to determine the mixed anomalies of $\cG\cong \Z_k$ with the global symmetries. Twisting by $\cG$ explicitly breaks the global symmetries as $SO(4)\to G_\text{global}\cong U(1)^2$. The following parameterisation of this unbroken symmetry will prove helpful,
\begin{align}
  G_\text{global} = \frac{U(1)_1 \times U(1)_2}{\Z_k}\,,\qquad (g_1,g_2)\sim (e^{2\pi iq/k}g_1,e^{-2\pi ip/k}g_2)\,,
  \label{eq: G1 and G2}
\end{align}
where the charges of $U(1)_1,U(1)_2\cong U(1)$ are taken to be
\begin{align}
    \begin{array}{c|cc}
        & \psi_{1} & \psi_{2} \\ \hline
        U(1)_1 & p & q \\
        U(1)_2 & q & -p \\
    \end{array}
\label{eq: G1 and G2 def}
\end{align}
It will be useful to sometimes talk about the charge matrix of $(\psi_1,\psi_2)$, which in this case is \begin{align}
  Q=\begin{pmatrix}
	p & q \\ q & -p
\end{pmatrix} \,.
\end{align}
The quotient in \eqref{eq: G1 and G2} is required to have the correct global form, since $(e^{2\pi iq/k},e^{-2\pi ip/k})\in U(1)_1\times U(1)_2$ is seen to act trivially on the fermion fields $\psi_1,\psi_2$.

Let us state explicitly the torus partition function of the original theory $\cT$ of 2 Weyl fermions, graded by fugacities for $U(1)_1$ and $U(1)_2$. There are four spin structures which we label by $(u,v)\in \Z_2\times \Z_2$ so that e.g.\ $(u,v)=(0,0)$ is NS-NS, $(1,0)$ is R-NS, etc. The partition function is then
\begin{align}
    Z^{(u,v)}(\tau,\mu_i) &= \text{Tr}_{\cH^u} (-1)^{vF}  \bar{\zeta}^{\bar{h}-1/24} z_1^{Q_1}z_2^{Q_2} \nn\\
    &\quad= \frac{1}{\overline{\eta(\tau)}^2} \sum_{n_1,n_2\in \Z-\frac{u}{2}} (-1)^{v(n_1+n_2)} \bar{\zeta}^{n_1^2/2 +n_2^2/2} z_1^{pn_1+qn_2}z_2^{qn_1 - pn_2 } \,,
\label{eq: original PF}
\end{align}
where $Q_1,Q_2$ are the Noether charges of $U(1)_1,U(1)_2$, respectively, and $\cH^0,\cH^1$ are the NS and R Hilbert spaces, respectively. The chemical potentials $\mu_i$ and fugacities $z_i$ are related as $z_i = e^{\mu_i}$. It will be useful later to note that $(-1)^F$ can be absorbed into the fugacities such that we have, for instance,\footnote{The shift by $2\pi i q$ in the second argument is trivial in the NS sector ($u=0$), but non-trivial and necessary in the R sector ($u=1$).}
\begin{align}
  Z^{(u,1)}(\tau,\mu_1,\mu_2) =  Z^{(u,0)}(\tau,\mu_1 + \pi i,\mu_2+\pi i (2q-1)) \,.
  \label{eq: sector shift T}
\end{align}
We now proceed to consider the mixed anomalies of $\cG$. It is straightforward to see that the mixed anomalies of $\cG$ with both $U(1)_1$ and $U(1)_2$ vanish. However $G_\text{global}$ does have a mixed anomaly with $\cG$, which can be diagnosed by considering the Hilbert space $\cH_{\alpha}^{0} $ of the theory on $S^1$ with NS boundary conditions augmented by a twist $\alpha=0,1,\dots,k-1$ in $\cG$, such that
\begin{align}
\psi_1(\sigma + 2\pi) = - e^{-2\pi i \alpha q/k}\psi_1(\sigma)\,,\qquad   \psi_2(\sigma + 2\pi) = - e^{2\pi i \alpha p/k}\psi_2(\sigma) \,.
\label{eq: twist}
\end{align}
Our conventions are such that this twist is enacted by the insertion of a topological line $\eta^\alpha$ pointing along the \emph{negative} time direction, where $\eta$ is the generator of $\cG$. Equivalently, by the operator-state map we are studying operators living at the end (rather than the beginning) of the line $\eta^\alpha$.

Further grading with fugacities for $U(1)_1$ and $U(1)_2$ yields
\begin{align}
  \text{Tr}_{\cH^{0}_\alpha} \,\bar{\zeta}^{\bar{h}-1/24} z_1^{Q_1} z_2^{Q_2} = \frac{1}{\overline{\eta(\tau)}^2} \sum_{n_1,n_2\in \Z} \bar{\zeta}^{(n_1+\alpha q/k)^2/2 +(n_2-\alpha p/k)^2/2} z_1^{pn_1+qn_2}z_2^{qn_1 - pn_2 + \alpha } \,.
\label{eq: NS PF}
\end{align}
Then, the mixed anomaly between $G_\text{global}$ and $\cG$ is diagnosed by the non-invariance of this expression under
\begin{align}
  (z_1,z_2) \to (e^{2\pi iq/k}z_1,e^{-2\pi ip/k}z_2)\,,
\end{align}
in the twisted sectors where $\alpha\neq 0$. The correct global form of the global symmetry once we include the twisted sectors is precisely the $k$-fold cover $U(1)_1\times U(1)_2$.\footnote{This mixed anomaly between $\cG$ and $G_\text{global}$ and the resulting $k$-fold cover can also be seen by noting that the $U(1) \subset G_\text{global}$ subgroup
\begin{align}\label{eq:anom_subgroup}
\{(e^{ i q\theta/k},e^{- i p\theta/k})\in (U(1)_1\times U(1)_2)/\Z_k : \theta\in [0,2\pi)\}\cong U(1) \,,
\end{align}
has a mixed anomaly $p$ mod $k$ with $\cG$, and $p$ is coprime to $k$.} This is what makes the parameterisation \eqref{eq: G1 and G2 def} of the Cartan subgroup $U(1)^2$ convenient. As in the case of the compact boson, this implies that there are $k$ choices of possible actions of $\cG$ in the NS twisted sectors, corresponding to the $k$ different ways of embedding $\cG$ into $U(1)_1\times U(1)_2$ in a manner compatible with the projection $U(1)_1\times U(1)_2\to G_{\text{global}}$.

However, as we are working with fermions, we need to look also at the Ramond sector, which introduces a further subtlety. In the $\cG$-twisted Ramond sectors, we have boundary conditions
\begin{align}
\psi_1(\sigma + 2\pi) =  e^{-2\pi i \alpha q/k}\psi_1(\sigma),\qquad   \psi_2(\sigma + 2\pi) =  e^{2\pi i \alpha p/k}\psi_2(\sigma) \,,
\end{align}
and find
\begin{align}
  \text{Tr}_{\cH^{1}_\alpha} \,\bar{\zeta}^{\bar{h}-1/24} z_1^{Q_1} z_2^{Q_2} = \frac{1}{\overline{\eta(\tau)}^2} \sum_{n_1,n_2\in \Z-\frac{1}{2}} \bar{\zeta}^{(n_1+\alpha q/k)^2/2 +(n_2-\alpha p/k)^2/2} z_1^{pn_1+qn_2}z_2^{qn_1 - pn_2 + \alpha } \,.
\end{align}
We see that in the Ramond sector the global symmetry is further fractionalised, with all states carrying charges $Q_1,Q_2\in \frac{1}{2}+\Z$. The product of double covers of $U(1)_1$ and $U(1)_2$, which we write
\begin{align}
    \widetilde{U(1)}_1\times \widetilde{U(1)}_2\,,
\end{align}
acts non-projectively but its action is not faithful since
\begin{align}
    e^{2\pi iQ_1+2\pi i Q_2}\in \widetilde{U(1)}_1\times \widetilde{U(1)}_2\,,
\end{align}
acts trivially on all sectors. Ultimately, the global symmetry which acts non-projectively and faithfully on all the $\cG$-twisted sectors, $(\cup_\alpha \cH_\alpha^{0})\cup (\cup_\alpha \cH_\alpha^{1})$, is given by
\begin{align}
  G_\text{faithful} &\cong \frac{\widetilde{U(1)}_1\times \widetilde{U(1)}_2}{\Z_2}\,\nn \\
  &=\{e^{2i\varphi_1Q_1 + 2i \varphi_2 Q_2 }: \varphi_1,\varphi_2\in [0,2\pi) \,,(\varphi_1,\varphi_2)\sim (\varphi_1+\pi,\varphi_2+\pi)\}\nn\\
  &= \{e^{i\theta(Q_1-Q_2) + 2i \phi Q_2 }: \theta,\phi\in [0,2\pi)\} \,.
\end{align}
This is what we need to proceed. The possible choices of the $\cG$-action on the twisted sectors correspond to the $\Z_k$ subgroups of $G_\text{faithful}$ which map to $\cG\subset G_\text{global}$ under the quotient map
\begin{align}
  G_\text{faithful} \to \frac{G_\text{faithful}}{\Z_{2k}} \cong G_\text{global}\,,
  \label{eq: group homo}
\end{align}
where the $\Z_{2k}$ quotient is generated by $e^{2\pi i(qQ_1 - pQ_2)/k}$. There are precisely $k$ such subgroups, which we can take to be generated by
\begin{align}\label{eq:twisting}
  \exp\left(\frac{2\pi i}{k}(Q_2 + t(qQ_1 - pQ_2))\right)\,,
\end{align}
where
\begin{align}\label{eq:t_range}
  t\in \{1,3,5,\dots,2k-1\}\,,
\end{align}
labels the $k$ distinct choices. A choice of $t$ thus fixes the $\cG$-action on all sectors.

It then remains to gauge $\cG$. We shall be interested in the torus partition function which, as the theory is fermionic, admits four spin structures: NS-NS, R-NS, NS-R and R-R. The latter two spin structures, with Ramond boundary conditions around the temporal cycle, require us to insert $(-1)^F$ in the Hilbert space picture. We must therefore fix an action of $(-1)^F$ on all sectors, which amounts to fixing a $\Z_2$ subgroup of $G_\text{faithful}$ with the property that it coincides with the known action of $(-1)^F$ on the $\cG$-untwisted NS and R Hilbert spaces, as can be read off from \eqref{eq: original PF}. There is a unique such subgroup, which is generated by
\begin{align}
  \exp \Big(\pi i \left(Q_1+(2q-1)Q_2\right)\Big) \,.
\end{align}
%
%
%
%
We then find that the torus partition function of the gauge theory with spin structure $(u,v)$ is given by
\begin{align}
    \hat{Z}^{(u,v)}_{t}(\tau,\mu_i) 	
    &= \frac{1}{k} \sum_{\alpha,\beta=0}^{k-1} \hat{Z}_{t,(\alpha,\beta)}^{(u,v)} (\tau,\mu_i)\,,
    \label{eq:Z_gauged}
\end{align}
where $\hat{Z}_{t,(\alpha,\beta)}^{(u,v)}(\tau,\mu_i) $ is the partition function in a sector of $\cG\cong \Z_{k}$ holonomies $\alpha,\beta\in \{0,1,\dots,k-1\}$. It is given explicitly by
 \begin{align}
    \hat{Z}_{t,(\alpha,\beta)}^{(u,v)}(\tau,\mu_i) = \frac{1}{\overline{\eta(\tau)}^2} \sum_{n_1,n_2\in \Z-\frac{u}{2}}& \bar{\zeta}^{(n_1+\alpha q/k)^2/2 +(n_2-\alpha p/k)^2/2} z_1^{pn_1+qn_2}z_2^{qn_1 - pn_2 + \alpha }		\nn\\
    & \times \exp\left(\frac{2\pi i \beta}{k} \big(qn_1 - pn_2+\alpha(1-tp)\big)- \pi i u\beta\right)		\nn\\
    & \times \exp\big(\pi i v \left(n_1+n_2+\alpha \right)\big)	\,.
\label{eq: twisted PF}
\end{align}
The first exponential is a result of introducing the twist by \eqref{eq:twisting} while the second comes from the insertion of $(-1)^{vF}$. Several simplifying manipulations have been made, for example using that $t$ and $p\pm q$ are odd.
As a consistency check, one can verify that this expression is indeed independently invariant under all of $u\to u+2,v\to v+2, \alpha\to \alpha+k,\beta\to\beta+k$, as it must be. 

Note that, like in the ungauged theory, we can absorb the insertion of $(-1)^F$ into the fugacities, for instance by
\begin{align}
  \hat{Z}_t^{(u,1)}(\tau,\mu_1,\mu_2) =  \hat{Z}_t^{(u,0)}(\tau,\mu_1 + \pi i,\mu_2+\pi i (2q-1)) \,.
\label{eq: sector shift T/G}
\end{align}

\subsection{Modular transformations}\label{sec:mixed_gravitational_anomaly}

So far we know that $\calG$ has no 't Hooft anomaly, and so can be consistently gauged. We now wish to understand the modular properties of the gauged theory. Recall that our aim is to find a $\calG$ such that the theory is mapped back to itself under gauging. For such a self-duality to be possible, the modular properties of the theory before and after gauging must be the same. Imposing this constraint will uniquely fix the value of $t$ in the choice of $\cG$-action \eqref{eq:twisting}.

One may intuitively think of this constraint as ensuring that there is no mixed $\cG$-gravitational anomaly, but this is imprecise.
For a unitary two-dimensional fermionic theory with a $\cG=\Z_k$ symmetry, the anomaly is valued in $\Z \times \Z_k$ when $k$ is odd. The $\Z$ factor describes the standard gravitational anomaly, but the remaining $\Z_k$ factor cannot be said to be either a pure $\cG$ anomaly or a mixed $\cG$-gravitational anomaly; in some sense it is both. The construction of the previous subsection ensures that this $\Z_k$-valued anomaly is not present in our case. The correct way to interpret this is that there must be at least one value of $t$ such that the resulting gauged theory is a unitary fermionic theory with the same gravitational anomaly as the original theory of two Weyl fermions. In fact, there is a \textit{unique} choice of $t$ with this property, which we will denote by $t_*$. This is because different such choices of $t$ would be related by stacking a fermionic $\cG$-SPT phase and for $k$ odd there are no non-trivial SPTs to stack.\footnote{The only allowed one would be $(-1)^{\text{Arf}(\rho)}$, where $\rho$ denotes the spin structure, but this does not become trivial when the $\cG$ background field is turned off. We will return to the role of $(-1)^{\text{Arf}(\rho)}$ in the gauging in section~\ref{subsec: self-duality}.}$^,$\footnote{One may ask what becomes of the theory if we use other values of $t\neq t_*$. The results of section~\ref{subsec: self-duality} will imply that for these values of $t$ the gauged theory is not fermionic. Rather, its primary operators have spins $h-\bar{h}\in \frac{1}{2k}\Z$. These `parafermionic' theories fall outside of the anomaly classification used above. We will avoid these cases in the present work, but highlight recent discussion of self-dualities and discrete gauging in similar theories in \cite{Thorngren:2021yso}.}

In order to proceed, we must determine $t_*$. This can be done by studying modular transformations of the gauged partition function.





In this subsection we turn off the fugacities $z_1=z_2=1$ for notational simplicity. 
We must ensure that under modular transformations of the torus the $\cG$-twisted partition function $Z_{t,(\alpha,\beta)}^{(u,v)}(\tau)$ transforms in the same way as the untwisted partition function.
To make this precise we have to take into account that modular transformations act non-trivially on the holonomies as\footnote{We reuse the symbol $\cT$ here, which also denotes the theory of 2 Weyl fermions. }
\begin{equation}
    \calT: (\alpha,\beta) \to (\alpha,\beta+\alpha) \qc \calS : (\alpha,\beta) \to (\beta, -\alpha) \,.
\end{equation}
We first consider the modular properties of the NS-NS partition function. The untwisted partition function $Z^{(0,0)}(\tau) \equiv \hat{Z}^{(0,0)}_{t,(0,0)}(\tau)$ has
\begin{align}
    Z^{(0,0)}(-1/\tau) &= Z^{(0,0)}(\tau) \,,\label{eq:ungauged_NSNS_S} \\ Z^{(0,0)}(\tau+1) &= e^{i\pi /6} Z^{(0,1)}(\tau) \,.\label{eq:ungauged_NSNS_T}
\end{align}
Let's look at $\cS$ first. 
We require
\begin{equation}
    \hat{Z}^{(0,0)}_{t,(\beta,-\alpha)} (-1/\tau) = \hat{Z}^{(0,0)}_{t,(\alpha,\beta)} (\tau) \,.
    \label{eq: S transf}
\end{equation}
We then compute using Poisson resummation,
\begin{equation}
    \hat{Z}^{(0,0)}_{t,(\beta,-\alpha)}(-1/\tau) = \frac{1}{\overline{\eta(\tau)}^2} \sum_{n_1,n_2\in \Z} \bar{\zeta}^{(n_1+\alpha q/k)^2/2 +(n_2-\alpha p/k)^2/2} \exp\left(\frac{2\pi i \beta}{k} \left(qn_1 - pn_2+\alpha tp\right)\right) \,.
\end{equation}
By direct comparison with \eqref{eq: twisted PF}, we see that \eqref{eq: S transf} is satisfied if and only if 
\begin{align}
  1-2tp = 0\quad \text{mod }k \,.
  \label{eq: t constraint}
\end{align}
Using the fact that $2p$ is coprime to $k$, it follows that there is a unique value $t=t_*$ of the parameter $t\in \{1,3,\dots,2k-1\}$ which satisfies this condition.

This confirms that if $t\neq t_*$, the gauged theory transforms differently to the original theory under modular transformations. It remains to verify that when $t=t_*$, all modular transformations of the gauged theory agree with those of the ungauged theory in all sectors. We first find that under $\cT$ the gauged partition function transforms as
\begin{equation}
    \hat{Z}^{(0,0)}_{t_*,(\alpha,\beta+\alpha)}(\tau+1) = e^{i\pi/6} \hat{Z}^{(0,1)}_{t_*,(\alpha,\beta)}(\tau)\,,
\end{equation}
which, after comparison with \eqref{eq:ungauged_NSNS_T} completes our check that the gauged partition function transforms in the correct way under modular transformations in the NS-NS sector.
The calculation for the other spin structures follow similarly and there are no surprises. That is, with $t=t_*$, we find
\begin{align}
    \hat{Z}^{(0,0)}_{t_*,(\beta,-\alpha)} (-1/\tau) &= \hat{Z}^{(0,0)}_{t_*,(\alpha,\beta)} (\tau) \qc &\hat{Z}^{(0,0)}_{t_*,(\alpha,\beta+\alpha)}(\tau+1) &= e^{i\pi/6} \hat{Z}^{(0,1)}_{t_*,(\alpha,\beta)}(\tau) \,, \nn\\
    \hat{Z}^{(0,1)}_{t_*,(\beta,-\alpha)}(-1/\tau) &= \hat{Z}^{(1,0)}_{t_*,(\alpha,\beta)}(\tau) \qc &\hat{Z}^{(0,1)}_{t_*,(\alpha,\beta+\alpha)}(\tau+1) &= e^{i\pi/6} \hat{Z}^{(0,0)}_{t_*,(\alpha,\beta)} (\tau) \,, \nn\\
    \hat{Z}^{(1,0)}_{t_*,(\beta,-\alpha)}(-1/\tau) &= \hat{Z}^{(0,1)}_{t_*,(\alpha,\beta)}(\tau) \qc &\hat{Z}^{(1,0)}_{t_*,(\alpha,\beta+\alpha)}(\tau+1) &= e^{-i\pi/3} \hat{Z}^{(1,0)}_{t_*,(\alpha,\beta)} (\tau) \,, \nn\\
    \hat{Z}^{(1,1)}_{t_*,(\beta,-\alpha)} (-1/\tau) &= -\hat{Z}^{(1,1)}_{t_*,(\alpha,\beta+\alpha)} (\tau) \qc &\hat{Z}^{(1,1)}_{t_*,(\alpha,\beta+\alpha)} (\tau+1) &= e^{-i\pi/3} \hat{Z}^{(1,1)}_{t_*,(\alpha,\beta)}(\tau) \,. \label{eq:modularTransfs}
\end{align}
These transformations are represented in Figure~\ref{fig:modularTransfs} and are precisely the same modular transformations as the ungauged partition function $Z^{(u,v)}(\tau) \equiv \hat{Z}^{(u,v)}_{t,(0,0)}(\tau)$ (see, e.g.\ \cite{BoyleSmith:2024qgx}).

\begin{center}
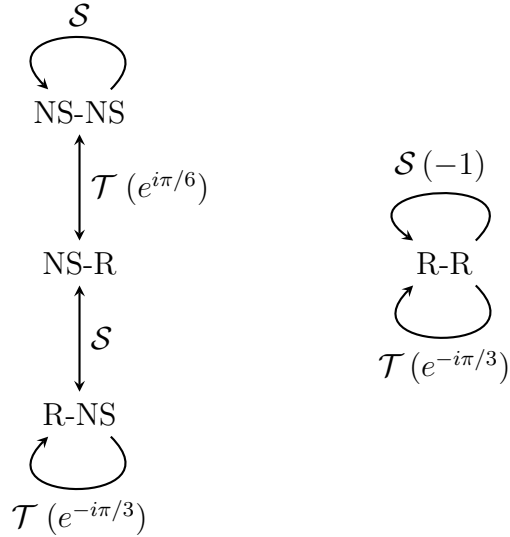

\begin{minipage}{0.8\textwidth}
	
    \centering

    \begin{tikzpicture}[
        >=stealth,
        thick,
        node distance=2cm,
        state/.style={minimum width=1.8cm, align=center}
    ]
    
    \node[state] (NSNS) {NS-NS};
    \node[state, below of=NSNS] (NSR) {NS-R};
    \node[state, below of=NSR] (RNS) {R-NS};
    
    \draw[<->] (NSNS.south) -- node[right] {$\calT\;(e^{i\pi/6})$} (NSR.north);
    \draw[<->] (NSR.south) -- node[right] {$\calS$} (RNS.north);
    
    \draw[->, looseness=4] ($(NSNS.north east)+(-0.5,0)$) [out=45, in=135] to node[above] {$\calS$} ($(NSNS.north west)+(0.5,0)$);
    
    \draw[->, looseness=4] ($(RNS.south east)+(-0.5,0)$) [out=-45,in=-135] to node[below] {$\calT\;(e^{-i\pi/3})$} ($(RNS.south west)+(0.5,0)$);
    
    \node[state, right=3cm of NSR] (RR) {R-R};
    
    \draw[->, looseness=4] ($(RR.north east)+(-0.5,0)$) [out=45,in=145] to node[above] {$\calS\,(-1)$} ($(RR.north west)+(0.5,0)$);
    
    \draw[->, looseness=4] ($(RR.south east)+(-0.5,0)$) [out=-45,in=-135] to node[below] {$\calT\,(e^{-i\pi/3})$} ($(RR.south west)+(0.5,0)$);
    
    \end{tikzpicture}\vspace{-0.05em}
    \captionof{figure}{Diagram of the modular $\calS,\calT$ transformations in \eqref{eq:modularTransfs}. Phases picked up under transformations are given in parentheses. See \cite{BoyleSmith:2024qgx} for the modular transformations of more general chiral CFTs.}
    \label{fig:modularTransfs}
\end{minipage}
\end{center}
We conclude that the gauged partition function transforms correctly under modular transformations if and only if $t=t_*$.

From one perspective, the choice of $t$ specifies the action of $\cG$ on the $\cG$-twisted sectors, as discussed around \eqref{eq: group homo}. From a Lagrangian perspective this corresponds to a specification of which counter-terms involving the $\cG$ background field are included in the gauging. Thus, there is a unique action of $\cG$ on the twisted sectors, or equivalently a unique choice of counter-terms, such that gauging $\cG$ does not affect the modular properties of the theory. This is a necessary condition for self-duality under gauging $\cG$.

\subsection{Demonstrating self-duality}\label{subsec: self-duality}

By gauging $\cG\cong \Z_k$ with the choice $t=t_*$ fixing the action of $\cG$ on the twisted sectors, we have shown that we land on a fermionic CFT which we denote $\cT/\cG$. For notational convenience, we henceforth fix $t=t_*$ and write $\hat{Z}^{(u,v)} \equiv \hat{Z}^{(u,v)}_t$ for the partition function of $\cT/\cG$ with spin structure $\rho=(u,v)$. The hat distinguishes this from the partition function $Z^{(u,v)}$ of the ungauged theory $\cT$ as given in \eqref{eq: original PF}.

As discussed above, the gauge theory $\cT/\cG$ is necessarily the same as the theory $\cT$ we started with. This should be contrasted with the compact boson example, whereby gauging can (and generically does) send us to another point on the $c=1$ bosonic conformal manifold. It will be important however to explicitly construct a map between $\cT/\cG$ and $\cT$, as we will do now.


%

\subsubsection*{The Neveu--Schwarz sector}

Let us first consider the theory $\cT/\cG$ with NS boundary conditions around the spatial circle. Since we have the relations \eqref{eq: sector shift T} and \eqref{eq: sector shift T/G}, it suffices to study the NS-NS partition function with fugacities  $z_1,z_2$ turned on for $U(1)_1,U(1)_2$, respectively. We find
\begin{align}
    \hat{Z}^{(0,0)}(\tau,\mu_i)&= \frac{1}{k}\sum_{\alpha,\beta=0}^{k-1}\frac{1}{\overline{\eta(\tau)}^2} \sum_{n_1,n_2\in \Z} \bar{\zeta}^{(n_1+\alpha q/k)^2/2 +(n_2-\alpha p/k)^2/2} z_1^{pn_1+qn_2}z_2^{qn_1 - pn_2 + \alpha }		\nn\\
    &\hspace{45mm} \times \exp\left(\frac{2\pi i \beta}{k} \left(qn_1 - pn_2+ t_*p\alpha\right)\right) \,.	
\label{eq: twisted PF again}
\end{align}
The sum over $\beta$ projects the other three sums onto those values of $(n_1,n_2,\alpha)$ satisfying
\begin{align}
  qn_1 - pn_2 +  t_* p \alpha = 0 \quad \mod k \,,
\label{eq: constraint}
\end{align}
which, since $k$ is odd, is equivalent to
\begin{align}
 \alpha - 2(pn_2-qn_1) = 0 \quad \mod k \,,
\label{eq: constraint better}
\end{align}
where we've used $2pt_*=1$ mod $k$.
The solutions $(n_1,n_2,\alpha)$ of this equation can be parameterised by a pair of integers $m_1(n_1,n_2,\alpha),m_2(n_1,n_2,\alpha)\in \Z$ defined by
\begin{align}\label{eq:iso_NSNS}
m_1(\alpha,n_1,n_2) &= n_1-
  \frac{q}{k}\big(  \alpha - 2(pn_2-q n_1) \big) \,,         \nn\\
m_2(\alpha,n_1,n_2) &= n_2+
  \frac{p}{k}\big( \alpha - 2(pn_2-q n_1)  \big) \,.
\end{align}
This can be written equivalently in the compact form
\begin{align}
  \begin{pmatrix}
  	m_1(\alpha,n_1,n_2) \\ m_2(\alpha,n_1,n_2)
  \end{pmatrix} = \frac{1}{k} \begin{pmatrix}
  	a & b \\ b & -a
  \end{pmatrix}\begin{pmatrix}
  	n_1 + \frac{\alpha q}{k} \\ n_2 - \frac{\alpha p}{k}
  \end{pmatrix}\,,
\label{eq: inverse map matrices}
\end{align}
with $a,b$ defined in \eqref{eq: Euclid}, although this obscures the fact that $m_1,m_2$ are indeed integers by virtue of the constraint \eqref{eq: constraint better}. Crucially, this is a one-to-one parameterisation, as evidenced by the inverse map
\begin{align}
\alpha(m_1,m_2)   &= 2(pm_2-qm_1)\quad (\text{mod }k), \,  \qquad \alpha(m_1,m_2)\in \{0,1,\dots,k-1\},      \nn\\
n_1 (m_1,m_2)     &= m_1 - \frac{q}{k}\big( \alpha(m_1,m_2) - 2(pm_2-q m_1) \big) \,,  \nn\\
n_2 (m_1,m_2)     &= m_2 + \frac{p}{k}\big( \alpha(m_1,m_2) - 2(pm_2-q m_1) \big) \,. \label{eq:inverse} 
\end{align}
The virtue of this parameterisation is that the scaling dimension of the operator $(n_1,n_2,\alpha)$ appearing in \eqref{eq: twisted PF again} becomes simply
\begin{align}
  \frac{1}{2}\left(n_1 + \frac{\alpha q}{k}\right)^2 + \frac{1}{2}\left(n_2 - \frac{\alpha p}{k}\right)^2 = \frac{1}{2}m_1^2 + \frac{1}{2}m_2^2\,,
  \label{eq: conformal}
\end{align}
recovering, as claimed, the spectrum of two Weyl fermions.

Furthermore, keeping track of chemical potentials, we find 
\begin{align}
  \hat{Z}^{(0,0)}(\tau,\mu_1,\mu_2) &= \frac{1}{\overline{\eta(\tau)}^2} \sum_{m_1,m_2\in \Z} \bar{\zeta}^{m_1^2/2 + m_2^2/2} z_1^{pm_1+qm_2}z_2^{-qm_1+pm_2}	\nn\\
  &= Z^{(0,0)}(\tau,\mu_1,-\mu_2) \,.
  \label{eq: NS-NS bad duality}
\end{align}
Note in particular the non-trivial action on the chemical potentials, and thus on the action of $U(1)_1\times U(1)_2$. Indeed, we have landed on the NS-NS partition function of two Weyl fermions $\psi_1',\psi_2'$ which carry the following charges under $U(1)_1\times U(1)_2$,
\begin{align}
    \begin{array}{c|cc}
        & \psi_{1}' & \psi_{2}' \\ \hline
        U(1)_1 & p & q \\
        U(1)_2 & -q & p \\
    \end{array}
\label{eq: G1 and G2 twisted}
\end{align}
to be contrasted with \eqref{eq: G1 and G2 def}. In particular, the $\Z_k\subset U(1)_1\times U(1)_2$ appearing in the quotient in \eqref{eq: G1 and G2} no longer acts trivially on \textit{local} operators of the free fermion realisation of the gauged theory. However, there is a different $\Z_k$ that does act trivially, generated by $(e^{2\pi iq/k},e^{2\pi ip/k})$. In other words, local operators of the gauged theory fractionalise the charges carried by local operators of the ungauged theory. This is not a surprise, and is just a simple consequence of the fact that certain twist operators of the ungauged theory become local operators of the gauged theory, and vice versa.

We also need to check that the action of $(-1)^F$ in the NS sector is preserved under the map \eqref{eq: inverse map matrices} between the operators of $\cT$ and $\cT/\cG$, which amounts to generalising \eqref{eq: NS-NS bad duality} to the NS-R sector. This is straightforward using \eqref{eq: sector shift T} and \eqref{eq: sector shift T/G} which imply
\begin{align}
   \hat{Z}^{(0,v)}(\tau,\mu_1,\mu_2) = Z^{(0,v)}(\tau,\mu_1,-\mu_2) \,.
   \label{eq: Z0v}
\end{align}
We have thus succeeded in mapping the NS Hilbert space of $\cT/\cG$ to that of $\cT$. 

This map is far from unique: for instance, after making the identification \eqref{eq: inverse map matrices} we could furthermore conjugate one of the fermions, e.g.\ $\psi_1\to \psi_1^\dagger$, corresponding to $m_1\to -m_1$, or else swap them $\psi_1\leftrightarrow \psi_2$, corresponding to $m_1\leftrightarrow m_2$. More generally, we can always compose our map \eqref{eq: inverse map matrices} with an invertible map from the NS sector of $\cT$ to itself. Such maps make up the $O(4)$ acting on four Majorana--Weyl fermions; swaps $\psi_1\leftrightarrow\psi_2$ and conjugations of each fermion generate a subgroup $D_4\subset O(4)$. Only elements of the subgroup $SO(4)\subset O(4)$ extend to automorphisms also of the Ramond sector, and thus of the theory $\cT$. This distinction will play an important role shortly.

\subsubsection*{The Ramond sector: Arf full or Arf empty?}

We next consider the Ramond sector. The calculations here are entirely analogous to the Neveu--Schwarz sector, and so we relegate details to Appendix~\ref{app: RR_duality}. We find that the map \eqref{eq: inverse map matrices} we have defined in the NS sector extends naturally to the Ramond sector. Using this map, we find
\begin{align}
 \hat{Z}^{(1,0)}(\tau,\mu_1,\mu_2) = Z^{(1,0)}(\tau, \mu_1,-\mu_2) \,.
\end{align}
This is not quite enough to conclude that the Ramond sectors of $\cT/\cG$ and $\cT$ are identified by this map, however. This is because in the Ramond sector of $\cT$, there is an ambiguity in how we define $(-1)^F$. The two consistent choices of $(-1)^F$ differ by $(-1)^F\to -(-1)^F$, and thus differ by a flip of sign of the R-R partition function. We have made a choice for $(-1)^F$ in our definition \eqref{eq: original PF} of the theory $\cT$, and need to check if our map recovers this choice or not. We compute the R-R partition function of the gauge theory, apply the Ramond sector extension of the map \eqref{eq: inverse map matrices}, and indeed find that the answer is no:
\begin{align}
 \hat{Z}^{(1,1)}(\tau,\mu_1,\mu_2) = -Z^{(1,1)}(\tau, \mu_1,-\mu_2) \,.
 \label{eq: RR sssssself duality}
\end{align}
Putting all four sectors together, we have
\begin{align}\label{eq:Arf_full_empty}
 \hat{Z}^{(u,v)}(\tau,\mu_1,\mu_2) = (-1)^{uv}Z^{(u,v)}(\tau,\mu_1,-\mu_2) = (-1)^{\text{Arf}(\rho)}Z^{(u,v)}(\tau,\mu_1,-\mu_2)\,,
\end{align}
where Arf$(\rho)$ is the Arf invariant, defined on a general Riemann surface with spin structure $\rho$ as
\begin{equation}
    \text{Arf}(\rho) = \begin{cases}
        0 & \rho \text{ even} \\
        1 & \rho \text{ odd}
    \end{cases}\,\,,
\end{equation}
We learn that, using the map \eqref{eq: inverse map matrices} to exchange $(n_1,n_2,\alpha)$ with $(m_1,m_2)$, we land not on the original theory $\cT$ but rather on $\cT\otimes \text{Arf}$. This is a problem for our intended self-duality construction, for which we must land back on the original theory $\cT$.

There is a quick fix available, however. Let us consider the slightly different map given by
\begin{align}
  \begin{pmatrix}
  	-m_2(\alpha,n_1,n_2) \\ m_1(\alpha,n_1,n_2)
  \end{pmatrix} = \frac{1}{k} \begin{pmatrix}
  	a & b \\ b & -a
  \end{pmatrix}\begin{pmatrix}
  	n_1 + \frac{\alpha q}{k} \\ n_2 - \frac{\alpha p}{k}
  \end{pmatrix} \,,
\label{eq: reference iso}
\end{align}
with inverse
\begin{align}
\alpha(m_1,m_2)   &= 2(pm_1+qm_2)\quad (\text{mod }k), \,  \qquad \alpha(m_1,m_2)\in \{0,1,\dots,k-1\},      \nn\\
n_1 (m_1,m_2)     &= -m_2 - \frac{q}{k}\big( \alpha(m_1,m_2) - 2(pm_1+qm_2) \big) \,,  \nn\\
n_2 (m_1,m_2)     &= m_1 + \frac{p}{k}\big( \alpha(m_1,m_2) - 2(pm_1+qm_2) \big) \,.
\label{eq: reference n and alpha}
\end{align}
In other words, we use \eqref{eq: inverse map matrices}, composed with a map $S':(\psi_1,\psi_2) \to (\psi_2^\dag,\psi_1)$. The fact that this operation absorbs the extra sign in \eqref{eq: RR sssssself duality} can be seen in an elementary way. Turning off the fugacities for brevity, the torus partition function of $\cT \otimes \text{Arf}$ is
\begin{align}
  (-1)^{uv} \frac{1}{\overline{\eta(\tau)}^2} \sum_{n_1,n_2\in \Z-\frac{u}{2}} (-1)^{v(n_1+n_2)} \bar{\zeta}^{n_1^2/2 +n_2^2/2}\,,
\end{align}
where, as always, we parameterise spin structures by $u,v\in \{0,1\}$. Then $S'$ enacts the change of variables $(n_1,n_2)\to (-n_2,n_1)$ under which this expression becomes
\begin{align}
 &(-1)^{uv}\frac{1}{\overline{\eta(\tau)}^2} \sum_{n_1,n_2\in \Z-\frac{u}{2}} (-1)^{v(-n_2+n_1)} \bar{\zeta}^{n_1^2/2 +n_2^2/2} 	\nn\\
 &\qquad = (-1)^{uv}\frac{1}{\overline{\eta(\tau)}^2} \sum_{n_1,n_2\in \Z-\frac{u}{2}} (-1)^{-2vn_2}(-1)^{v(n_1+n_2)} \bar{\zeta}^{n_1^2/2 +n_2^2/2}\nn\\
 &\qquad = \frac{1}{\overline{\eta(\tau)}^2} \sum_{n_1,n_2\in \Z-\frac{u}{2}} (-1)^{v(n_1+n_2)} \bar{\zeta}^{n_1^2/2 +n_2^2/2}\,, \label{eq:Arf_absorb}
\end{align}
which yields the partition function of $\cT$, as claimed.

Turning fugacities back on, we find that the map \eqref{eq: reference iso} yields
\begin{equation}\label{eq: reference_duality}
	\hat{Z}^{(u,v)}(\tau,\mu_1,\mu_2) = Z^{(u,v)}(\tau,\mu_2,\mu_1) \,,
\end{equation}
and thus we recover the original theory $\cT$, with charges under $U(1)_1$ and $U(1)_2$ swapped.

We say then that the map \eqref{eq: reference iso} defines an \emph{isomorphism} between $\cT/\cG$ and $\cT$. In general, we define an isomorphism between two QFTs as a bijection between the operators, both local and extended, of one theory and another. QFTs related by such an isomorphism are usually said to be dual. For two-dimensional CFTs, this takes the form of an invertible linear map between the Virasoro primaries of the two theories that preserves conformal weights $(h,\bar{h})$ and OPE coefficients.

\subsubsection*{Other isomorphisms}

Our choice of map \eqref{eq: reference iso} is not unique, and could be composed with any invertible line of $\cT$; that is, with any element of $SO(4)$. For example, we could consider
\begin{align}\label{eq:action_C}
  C\begin{pmatrix}
  	m_1(\alpha,n_1,n_2) \\ m_2(\alpha,n_1,n_2)
  \end{pmatrix} = \frac{1}{k} \begin{pmatrix}
  	a & b \\ b & -a
  \end{pmatrix}\begin{pmatrix}
  	n_1 + \frac{\alpha q}{k} \\ n_2 - \frac{\alpha p}{k}
  \end{pmatrix}\,,
\end{align}
for any of the matrices
\begin{align}\label{eq:four_Cs}
 C\in \left\{
 \pm \begin{pmatrix}
0 & -1 \\
1 & 0
\end{pmatrix},
\pm \begin{pmatrix}
1 & 0 \\
0 & -1
\end{pmatrix}
\right\}\,,
\end{align}
and similar manipulations to \eqref{eq:Arf_absorb} would go through. Indeed, the first of these is simply the choice \eqref{eq: reference iso} used above. It is straightforward to check these four maps are indeed related by composition with a suitable $SO(4)$ transformation. These four choices of map $\cT/\cG \to \cT$ are singled out from all others by the fact that they map the highest-weight representations of a chosen $U(1)^2$ current algebra of $\cT/\cG$, labelled by $(n_1,n_2,\alpha)$, to that of $\cT$, labelled by $(m_1,m_2)$. In other words, they are an artefact of our choice to work with two complex fermions, rather than four real ones.

Different choices of $C$ give different maps from $\cT/\cG \to \cT$ in \eqref{eq:action_C}, mapping the $U(1)_1\times U(1)_2$ fugacities differently. The generalisation of \eqref{eq: reference_duality} for a more general choice of $C$ is then (see Appendix~\ref{app: RR_duality} for the proof)
\begin{align}\label{eq:general duality}
	\hat{Z}^{(u,v)}(\tau,\mu_i) = Z^{(u,v)}(\tau,K_{ij} \mu_j) \,,
\end{align}
where $K$ is an orthogonal matrix given by
\begin{align}\label{eq:four K's}
    K \in \left\{ \pm \begin{pmatrix}
  	0 & 1 \\ 1 & 0
  \end{pmatrix} , \pm \frac{1}{k} \begin{pmatrix}
  	a & -b \\ b & a
  \end{pmatrix} \right\}\,,
\end{align}
for the choices of $C$ in \eqref{eq:four_Cs}, with $a,b$ given in \eqref{eq: Euclid}. Indeed, for the first choice of $C$, we recover \eqref{eq: reference_duality}.

As a side remark, we note that these considerations make manifest that the invertible 0-form symmetry of $\cT$ is $SO(4)$, not $O(4)$. Indeed, if we consider the $O(4)$ matrix acting on four Majorana--Weyl fermions which corresponds to the $C$ that yields \eqref{eq: reference iso}, we find (see Appendix~\ref{app: RR_duality})
\begin{align}
  C = \begin{pmatrix}
0 & -1 \\
1 & 0
\end{pmatrix} \quad \longrightarrow \quad \det\!{}_{O(4)}(C) = \det\begin{pmatrix}
0 & \sigma_3 \\
\mathds{1}_2 & 0
\end{pmatrix} = -1 \,,
\end{align}
so it is not an element of $SO(4)$. Those transformations which are elements of $SO(4)$ are symmetries of $\cT$, i.e.\ isomorphisms $\cT\to\cT$, while those which are elements of $O(4)$ but not $SO(4)$ are \emph{not} symmetries of $\cT$ and give isomorphisms $\cT \otimes \text{Arf} \to \cT$. The map $S':(\psi_1,\psi_2) \to (\psi_2^\dag, \psi_1)$ to which this $C$ corresponds is an example of the latter, as are all of the transformations in \eqref{eq:four_Cs}. This is why they can be used to rectify the fact that the isomorphism \eqref{eq: inverse map matrices} which we studied originally maps into $\cT\otimes \text{Arf}$.

\section{Non-invertible symmetries}\label{sec: Non-invertible interfaces and boundaries}

We have now established that the theory of two Weyl fermions is self-dual under gauging any anomaly-free symmetry. The aim of this section is to use these self-dualities to construct and study non-invertible topological defects of the theory.

We first review the half-space gauging construction of these defects in section~\ref{subsec:half_space_gauging}. In section~\ref{subsec:properties_defects_TYness} we will apply this to our setup, and study how extended operators are acted upon by the non-invertible lines. We then discuss in section~\ref{subsec:scattering_through} how local excitations scatter through them. Finally in section~\ref{sec:fusion_rules} we study some aspects of their fusion.

\subsection{A review of half-space gauging}\label{subsec:half_space_gauging}

Whenever a theory $\cT$ is self-dual under the gauging of some discrete symmetry $\cG$, one can construct a corresponding non-invertible codimension-one topological defect in the theory. The procedure to do so is called \emph{half-space gauging}, and the resulting non-invertible defects, often called \emph{self-duality defects}, have been studied in a number of contexts in both two \cite{Frohlich:2004ef,Frohlich:2006ch,Fuchs:2007tx,Bachas:2012bj,Bachas:2013ora,Thorngren:2021yso,ChoiCordovaHsinLam2021,Damia2024,Argurio:2024ewp,Niro:2022ctq,Bharadwaj:2024gpj,Arias-Tamargo:2025xdd,Arias-Tamargo:2025fhv,Arias-Tamargo:2025atr,Pace:2024oys,Pace:2024tgk} and higher dimensions \cite{Kaidi:2021xfk,Bashmakov:2022uek,Antinucci:2022cdi,Carta:2023bqn}. We now review this construction.

One starts by splitting spacetime into a left and right half-space, and gauging $\cG$ only in, say, the right half-space. This requires a choice of how to gauge $\cG$, which is not generically unique. It also requires a choice of boundary conditions for the $\cG$ gauge field at the interface; Dirichlet boundary conditions are always available, which render the interface topological.

At this point, one has two different theories joined at the interface. Suppose, however, that the theory is self-dual under gauging $\cG$. This means that there exists an isomorphism $D$,
\begin{align}
  \cT/\cG \xlongrightarrow{D} \cT \,.
\end{align}
We can think of $D$ as defining a topological interface between $\cT/\cG$ and $\cT$ which, by a slight abuse of notation, we will also refer to as $D$. One can thus consider the setup in Figure~\ref{fig: half-space setup}. We can use the topological nature of both $\bra{\text{Dir}}$ and $D$ to stack them on top of one another. In doing so, we define a topological line of $\cT$, that we will call $\cN$.
\begin{center}
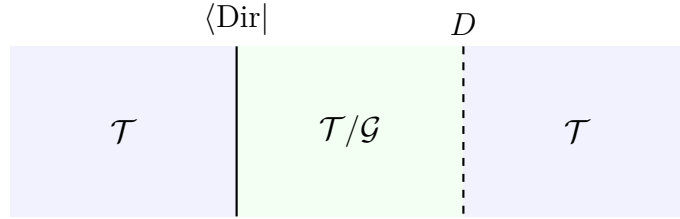
\vspace{1em}
\begin{minipage}{0.8\textwidth}
\[
    \begin{tikzpicture}[scale=0.75]
        \fill[blue!5] (0,0) rectangle (4,-3);
        \fill[green!5] (4,0) rectangle (8,-3);
        \fill[blue!5] (8,0) rectangle (12,-3);
        \draw[thick] (4,0) -- (4,-3);
        \draw[thick,dashed] (8,0) -- (8,-3);
        \node at (2,-1.5) {$\cT$};
        \node[above] at (4,0) {$\bra{\text{Dir}}$};
        \node at (6,-1.5) {$\cT/\cG $};
        \node[above] at (8,0) {$D$};
        \node at (10,-1.5) {$\cT$};
    \end{tikzpicture}\vspace{-1em}
\]
    \captionof{figure}{Construction of the self-duality defect $\cN$.} \label{fig: half-space setup}
\end{minipage}\vspace{1em}
\end{center}
As previously discussed, given an isomorphism $D$ between $\cT/\cG$ and $\cT$, we can always construct another isomorphism $D'$ by composition with some isomorphism from $\cT$ to itself:
\begin{align}
  \cT/\cG \xrightarrow{\,\,\,D'\,\,\,} T \qquad \equiv \qquad \cT/\cG \xrightarrow{\,\,\,D\,\,\,} \cT \xrightarrow{\,\,\,U\,\,\,} \cT \,.
\end{align}
The automorphisms $U$ are nothing but the invertible global symmetries of $\cT$. This implies the fusion rule
\begin{align}\label{eq:D' from D}
  D' = D \times U\,,
\end{align}
for the interfaces $D,D'$. It is clear that the line $\cN'$ corresponding to the choice of isomorphism $D'$ is related to $\cN$ by fusion with the invertible line $U$:
 \begin{align}\label{eq:N' from N}
  \cN' = \cN \times U \,.
\end{align}
Let us now specialise to $\cT$ the theory of two Weyl fermions, and consider gauging $\cG = \cG_{(p,q)}$ in a half-space for some particular $p,q$ coprime with $pq$ even. We know that there is a unique way to gauge $\cG$, and so all we need now is a choice of isomorphism $D$ between $\cT/\cG$ and $\cT$. There is no canonical choice of this isomorphism. However, safe in the knowledge that the resulting non-invertible defect $\cN$ can be turned into the defect we would have found by choosing $D'$ by fusion with some invertible line $U\in SO(4)$, we will choose a reference isomorphism $D$ given by \eqref{eq: reference iso}.\footnote{Given this choice of $D$, the equivalences (\ref{eq: group equivalences}) imply that the corresponding lines $\cN_{(p,q)}$ are equal up to fusion with invertible lines in $SO(4)$. Indeed, we find $\cN_{(p,q)}=\cN_{(-p,-q)}= \cN_{(q,-p)}\tilde{C} = \cN_{(-q,p)}\tilde{C}$ where $\tilde{C}$ is the $SO(4)$ line enacting charge conjugation $(\psi_1,\psi_2)\to (\psi_1^\dagger,\psi_2^\dagger)$.}

\subsection{Wilson lines and twist operators}\label{subsec:properties_defects_TYness}

We next want to understand how topological lines are mapped under the isomorphism $D$ corresponding to our line $\cN$, which will then be important for the scattering through it.

The gauge theory $\cT/\cG$ has an anomaly-free symmetry $\widehat{\cG} \cong \Z_k$, generated by the topological Wilson lines $W^a$, $a=0,1,\dots,k-1$, of the $\cG$ gauge field. This is known as the \emph{dual symmetry} (or \emph{quantum symmetry}) of $\cG$. The topological Wilson lines of $\cT/\cG$ must be mapped by $D$ to a $\Z_k$ subgroup of the invertible lines of $\cT$, which we denote $\cG'=D(\widehat{\cG})$. We note that $\cG'$ is generically not the same as $\cG$, while both are isomorphic to $\Z_k$ as groups. A generator of $\cG'$, which we denote $\eta'$, can be determined by finding which line in $\cT$ corresponds to $W$ in $\cT/\cG$.

We will now show that, for the isomorphism $D$ given in \eqref{eq: reference iso}, we indeed have $\cG'=\cG$.
For this to be the case, we must have
\begin{align}
  \langle \eta  \rangle_\cT  = \langle W^r \rangle_{\cT/\cG} \,,
  \label{eq: W eta new}
\end{align}
for some $r$ coprime to $k$. 
It will be notationally convenient to consider $\langle W^p\rangle_{\cT/\cG}$ which also generates $\widehat{\cG}$ since $p$ is coprime to $k$. Choosing the Wilson line to wrap the spatial circle, we have the NS-NS partition function\footnote{The sign appearing in the Wilson line contribution $-2\pi i\alpha/k$ can be traced back to our choice \eqref{eq: twist} of twist convention.}
\begin{align}
   \langle W^p\rangle_{\cT/\cG} = \sum_{\alpha,\beta=0}^{k-1}&\frac{1}{\overline{\eta(\tau)}^2} \sum_{n_1,n_2\in \Z} \bar{\zeta}^{(n_1+\alpha q/k)^2/2 +(n_2-\alpha p/k)^2/2} 	\nn\\
   &\times  \exp\left(-\frac{2\pi ip\alpha}{k}+\frac{2\pi i \beta}{k} \left(qn_1 - pn_2+\alpha(1-t_*p)\right)\right)\,.
\label{eq: Wp exp}
\end{align}
Meanwhile the expectation value $\langle \eta^\gamma \rangle_\cT $ where $\eta^\gamma$ is wrapped on the spatial cycle and $\gamma\in \{0,\dots,k-1\}$ is
\begin{align}
    \langle\eta^\gamma\rangle_\cT = \frac{1}{\overline{\eta(\tau)}^2} \sum_{m_1,m_2\in \Z} \bar{\zeta}^{m_1^2/2 +m_2^2/2} \exp\left(\frac{2\pi i \gamma}{k} \left(q m_1 - p m_2\right)\right)\,,
\end{align}
by the definition of the symmetry $\cG$. Our aim is to show that $\langle W^p\rangle_{\cT/\cG}$ can be brought to this form after application of the isomorphism $D$, for some value of $\gamma$.

In \eqref{eq: Wp exp} the constraint imposed by the sum over $\beta$ is solved as usual by replacing $(n_1,n_2,\alpha)$ by $(m_1,m_2)$ as fixed by \eqref{eq: reference iso}, which yields
\begin{equation}
    \alpha = 2(pm_1 + qm_2) \,,
\end{equation}
and thus
\begin{align}\label{eq: W eta relation}
   \langle W^p\rangle_{\cT/\cG} = \frac{1}{\overline{\eta(\tau)}^2} \sum_{m_1,m_2\in \Z} \bar{\zeta}^{m_1^2/2 +m_2^2/2} \exp\left(- \frac{4\pi i p}{k} \left(pm_1 + qm_2\right)\right) = \langle \eta^{2q} \rangle_{\cT} \,,
\end{align}
where we have used that $k=p^2+q^2$.
We thus confirm that \eqref{eq: W eta new} is satisfied with $r$ given by
\begin{align}
  r = (2q)^{-1} p \mod k \,.
  \label{eq: r fixed}
\end{align}
The value of $r$ fixes the relation between background fields for the dual symmetry $\widehat{\cG}$ of $\cT/\cG$ to those for $\cG$ in $\cT$.

\subsection{Scattering through self-duality defects}\label{subsec:scattering_through}

We can now consider what happens when excitations of $\cT$ scatter through the non-invertible defects $\cN$.

We already know what happens when generic operators pass through invertible defects, be they local operators or, more generally, twist operators living at the end of some other invertible line. This is fixed by the action of these invertible lines on local operators, along with their commutation relations (i.e.\ fusion) amongst each other. It thus suffices to understand what happens when excitations pass through $\cN$ defined with respect to the reference isomorphism \eqref{eq: reference iso}.

The map \eqref{eq: reference iso} provides a map between the local operators of $\cT/\cG$ and $\cT$. By virtue of \eqref{eq: W eta new}, this can be extended to a map between gauge non-invariant operators of $\cT/\cG$ (which live at the end of Wilson lines in $\cT/\cG$) and certain twist operators of $\cT$ (those living at the end of the $\cG$ lines in $\cT$). In detail, it follows from \eqref{eq: W eta new} that operators in $\cT/\cG$ living at the end of $W^g$ (i.e.\ those with gauge charge $-g$) must be in one-to-one correspondence with twist operators of $\cT$ living at the end of the topological line $\eta^{2qp^{-1}g}$, generalising the case $g=0$ given by (\ref{eq: reference iso}). In particular, it is useful to note that under this correspondence the $\cT/\cG$ operator
\begin{center}
\begin{tikzpicture}
\draw[
  thick,
  decorate,
  decoration={snake, amplitude=1mm, segment length=5mm}
] (0,0) -- (4,0);

\draw[thick, -{Triangle[scale=1.5]}]
(2.0,0) -- (2.15,0);

\node at (2,0.5) {$W^{pm_2 - qm_1}$};

\fill (4,0) circle (0.2);

\node[right] at (4.2,0) {$\psi_1^{m_1}\psi_2^{m_2}$};
\end{tikzpicture}
\end{center}
which, assuming the Wilson line has somewhere else to end, is gauge-invariant, corresponds to the following twist operator in $\cT$ under $D$:
\begin{center}
\begin{tikzpicture}
\draw[
  thick,
  postaction={decorate},
  decoration={
    markings,
    mark=at position 0.5 with {\arrow{Triangle[scale=1.5]}}
  }
]
(0,0) -- (4,0);

\node at (2,0.5) {$\eta^{\alpha(m_1,m_2)}$};

\fill (4,0) circle (0.2);

\node[right] at (4.2,0) {$\psi_1^{n_1(m_1,m_2)}\psi_2^{n_2(m_1,m_2)}$};
\end{tikzpicture}
\end{center}
Note that this is really a shorthand for the point operator corresponding to the state $(n_1,n_2)$ in $\cH_\alpha^\text{NS}$, as counted in the partition function (\ref{eq: NS PF}). From (\ref{eq: W eta new}) it follows that we must have
\begin{align}
\alpha(m_1,m_2)     &= (2qp^{-1})(pm_2 - qm_1) \mod k        \nn\\
                    &= 2(pm_1+qm_2) \mod k \,.
\end{align}
The values of $n_1(m_1,m_2),n_2(m_1,m_2)$ are then fixed by requiring that both operators have the same charges under $U(1)_1\times U(1)_2$, taking care of the fact that the fundamental fermion fields of each presentation differ by the swapping $U(1)_1\leftrightarrow U(1)_2$. These values are given in (\ref{eq: reference n and alpha}). Both operators then carry charge $(pm_1+qm_2,qm_1-pm_2)$ under $U(1)_1\times U(1)_2$.

Scattering a local excitation $\psi_1^{m_1}\psi_2^{m_2}$ through $\cN$ then follows straightforwardly from the construction of $\cN$ as depicted in Figure~\ref{fig: half-space setup}. Upon passing through the Dirichlet boundary $\bra{\text{Dir}}$, the operator first becomes $W^{pm_2-qm_1} \psi_1^{m_1}\psi_2^{m_2}$ with the Wilson line remaining attached to the Dirichlet boundary condition $\bra{\text{Dir}}$ where such lines can end. We can understand why this must be the case from the fact that $/\cG$ is a \emph{discrete} gauging. Such a gauging introduces no local dynamics or interactions, so the dynamics of $\cT$ (namely, $\psi_1^{m_1}\psi_2^{m_2} \rightarrow \psi_1^{m_1}\psi_2^{m_2}$) is unchanged, except for the addition of any necessary topological lines to ensure gauge invariance. Upon passing through $D$, it is mapped as above to the twist operator $(\alpha,n_1,n_2)$ of $\cT$ with $\alpha,n_1,n_2$ fixed by \eqref{eq: reference n and alpha}. This is depicted in Figure~\ref{fig: half-space scatter}.
\begin{center}
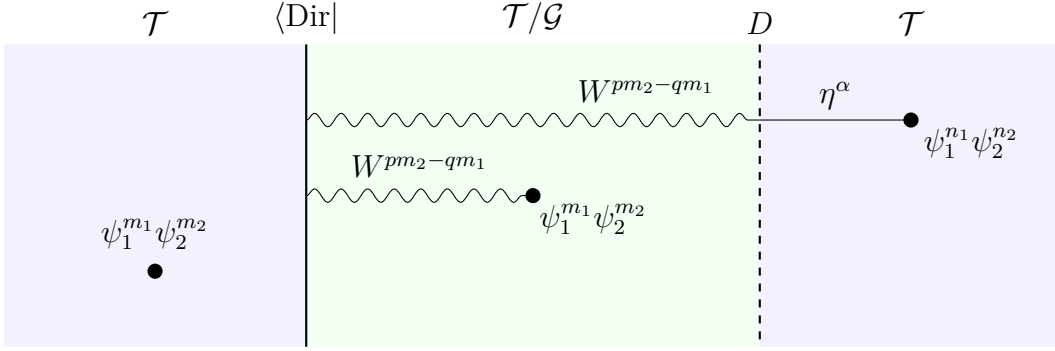

\begin{minipage}{\textwidth}\centering
\begin{minipage}{\textwidth}
\[
    \begin{tikzpicture}[baseline=-2.3cm]
        \fill[blue!5] (0,0) rectangle (4,-4);
        \fill[green!5] (4,0) rectangle (10,-4);
        \fill[blue!5] (10,0) rectangle (14,-4);
        \draw[thick] (4,0) -- (4,-4);
        \draw[thick,dashed] (10,0) -- (10,-4);
        \node[above] at (2,0) {$\cT$};
        \node[above] at (4,0) {$\bra{\text{Dir}}$};
        \node[above] at (7,0) {$\cT/\cG$};
        \node[above] at (10,0) {$D$};
        \node[above] at (12,0) {$\cT$};
        \fill (2,-1-2) circle (0.1);
        \fill (7,-2) circle (0.1);
        \fill (12,-3+2) circle (0.1);
        \draw[decorate, decoration=snake] (4,-2) -- (7,-2);
        \draw[decorate, decoration=snake] (4,-3+2) -- (10,-3+2);
        \draw (10,-3+2) -- (12,-3+2);
        \node[above] at (5.5,-1.9) {$W^{pm_2-qm_1}$};
        \node[above] at (8.5,-0.9) {$W^{pm_2-qm_1}$};
        \node[above] at (11,-3+2) {$\eta^\alpha $};
        \node[above] at (2,-0.85-2) {$\psi_1^{m_1}\psi_2^{m_2}$};
        \node[above] at (7.8,-1.95-0.7) {$\psi_1^{m_1}\psi_2^{m_2}$};
        \node[above] at (12.8,-2.95+2-0.7) {$\psi_1^{n_1}\psi_2^{n_2}$};
    \end{tikzpicture}
\]
\end{minipage}\vspace{1em}
\begin{minipage}{0.8\textwidth}
    \captionof{figure}{Scattering through $\cN$, decomposed into two stages. Here and throughout, time goes up.} \label{fig: half-space scatter}
\end{minipage}
\end{minipage}
\end{center}
We can finally fuse together the two interfaces to find the scattering through $\cN$ depicted in Figure~\ref{fig: N scattering}.
\begin{center}
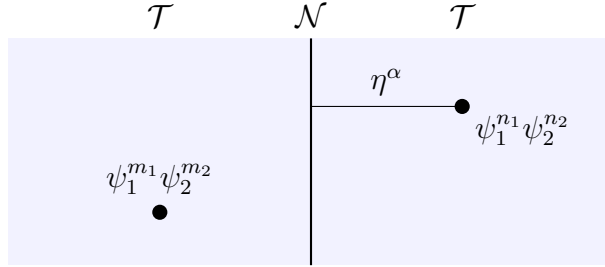
\vspace{1em}
\begin{minipage}{0.8\textwidth}
\[
    \begin{tikzpicture}
        \fill[blue!5] (2,0) rectangle (6,-3);
        \fill[blue!5] (6,0) rectangle (10,-3);
        \draw[thick] (6,0) -- (6,-3);
        \node[above] at (4,0) {$\cT$ };
        \node[above] at (6,0) {$\cN$};
        \node[above] at (8,0) {$\cT$};
        \fill (4,-1.2-1.6+0.5) circle (0.1);
        \fill (8,-2.8+1.6+0.3) circle (0.1);
        \draw (6,-2.8+1.6+0.3) -- (8,-2.8+1.6+0.3);
        \node[above] at (7,-2.8+1.6+0.3) {$\eta^{\alpha}$};
        \node[above] at (4,-1.1-1.6+0.5) {$\psi_1^{m_1}\psi_2^{m_2}$};
        \node[above] at (8.8,-1.85+0.3) {$\psi_1^{n_1}\psi_2^{n_2}$};
    \end{tikzpicture}\vspace{-1em}
\]
    \captionof{figure}{Scattering through $\cN $} \label{fig: N scattering}
\end{minipage}\vspace{1em}
\end{center}
We can ask then which global symmetries are preserved by this scattering, which is evident from the way in which the self-duality exchanges $U(1)_1\leftrightarrow U(1)_2$. One way to think about this is to say that the scattering process preserves a $U(1)_1'\times U(1)_2'$ that acts differently on the fundamental fields on each side of $\cN$. Namely, we can take it to act as
\begin{align}
    x<0:\quad \begin{array}{c|cc}
        & \psi_{1} & \psi_{2} \\ \hline
        U(1)_1' & p & q \\
        U(1)_2' & q & -p \\
    \end{array}\hspace{20mm}
    x>0:\quad \begin{array}{c|cc}
        & \psi_{1} & \psi_{2} \\ \hline
        U(1)_1' & q & -p \\
        U(1)_2' & p & q \\
    \end{array}
    \label{eq: non-local U(1) action}
\end{align}
Under this symmetry, both the ingoing and outgoing states appearing in Figure~\ref{fig: N scattering} carry charges $(pm_1+qm_2,qm_1-pm_2)$. The fact that the preserved symmetry acts differently on the left and right of $\cN$ is standard for non-invertible defects, and in fact it has been argued to be generic in two dimensions \cite{Prembabu:2025qvi}.

Equivalently, if we let $U[\theta_1,\theta_2]$ be the topological line of $\cT$ that enacts $(\psi_1,\psi_2)\to (e^{ip\theta_1 + i q \theta_2}\psi_1,e^{iq\theta_1-ip\theta_2}\psi_2)$, then we have the following relation between left and right fusion with $\cN$,
\begin{align}
U[\theta_1,\theta_2] \times \cN = \cN \times U[\theta_2,\theta_1] \,.
\end{align}
This can be seen by encircling the operator $\psi_1^{m_1}\psi_2^{m_2}$ with $U[\theta_1,\theta_2]$, and shrinking this circle on the operator either before or after pulling it through $\cN$.

\subsubsection*{Fixing phases}

Strictly speaking the map \eqref{eq: reference iso} only tells us how to relate the primaries of $\cT/\cG$ and $\cT$ up to phases. To fully specify the isomorphism $D$, and thus the defect $\cN$, we need to fix these phases. This can be done in a simple way.

%
Consider the pair of local operators
\begin{equation}\label{eq:neutrals}
    \Psi_1 = \psi_1^p \psi_2^q \qc \Psi_2 = \psi_1^q \psi_2^{-p} \,.
\end{equation}
They form a basis of the $\cG$-neutral primaries in $\cT$. Therefore, as they are scattered through $\cN$ they remain local; that is, they do not get attached to a line. Indeed, \eqref{eq: reference n and alpha} gives $\alpha=0$ for both of them. Moreover, the operators \eqref{eq:neutrals} are scattered in a particularly simple way by $\cN$. Under the map \eqref{eq: reference n and alpha}, we find that these two operators are exchanged, since $(m_1,m_2) = (p,q) \to (n_1,n_2) = (q,-p)$ and $(m_1,m_2) = (q, -p) \to (n_1,n_2) = (p, q)$. We can then fix the isomorphism $D$ and thus scattering through $\cN$ by imposing that these two operators do not acquire any additional phases:
\begin{equation}\label{eq:neutral_scat}
    \Psi_1 \to \Psi_2 \qc \Psi_2 \to \Psi_1 \,.
\end{equation}

\subsection{Fusion rules}\label{sec:fusion_rules}

On general grounds, any self-duality defect built from half-space gauging is non-invertible and will satisfy the following fusion rules \cite{Thorngren:2021yso,ChoiCordovaHsinLam2021,Kaidi:2021xfk}
\begin{align} \label{eq:almostTY}
    \eta^k=1\qc \eta\times\cN=\cN\qc \cN\times\overline{\cN}=\sum_{n=1}^k \eta^n\,,
\end{align}
where $\eta$ is the topological line generating $\cG$. The first fusion rule simply reflects the fact that $\cG\cong\Z_k$, while the second follows from the fact that $\cG$ is gauged on the right of $\cN$ so the generating line $\eta$ is trivial in the gauged theory. Note that the final expression in \eqref{eq:almostTY} features the orientation reversal $\overline{\cN}$ of $\cN$ which, from Figure~\ref{fig: half-space setup}, corresponds to the configuration shown in Figure~\ref{fig: orientation rev}.
\begin{center}
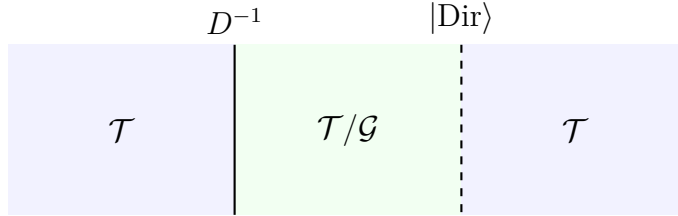
\vspace{1em}
\begin{minipage}{0.8\textwidth}
\[
    \begin{tikzpicture}[scale=0.75]
        \fill[blue!5] (0,0) rectangle (4,-3);
        \fill[green!5] (4,0) rectangle (8,-3);
        \fill[blue!5] (8,0) rectangle (12,-3);
        \draw[thick] (4,0) -- (4,-3);
        \draw[thick,dashed] (8,0) -- (8,-3);
        \node at (2,-1.5) {$\cT$};
        \node[above] at (4,0) {$D^{-1}$};
        \node at (6,-1.5) {$\cT/\cG $};
        \node[above] at (8,0) {$\ket{\text{Dir}}$};
        \node at (10,-1.5) {$\cT$};
    \end{tikzpicture}\vspace{-1em}
\]
    \captionof{figure}{The orientation reversal, $\overline{\cN}$, of the self-duality defect $\cN$.} \label{fig: orientation rev}
\end{minipage}\vspace{1em}
\end{center}
The final fusion rule in \eqref{eq:almostTY} is found immediately by composing Figures~\ref{fig: half-space setup} and \ref{fig: orientation rev}: the interfaces $D^{-1}$ and $D$ fuse to give the identity, leaving $\cT/\cG$ in a slab with Dirichlet boundary conditions on both sides. Moving the two boundaries together leaves a condensation of $\eta$. 

It also follows from \eqref{eq:almostTY} that $\overline{\cN}$ can absorb $\eta$ from the right:
\begin{equation}\label{eq: Nbar absorb eta}
    \overline{\cN} \times \eta = \overline{\cN} \,.
\end{equation}
Furthermore, we note that $\cN$ can absorb $\eta'$ from the right-hand side,
\begin{equation}\label{eq:eta' absorb}
    \cN \times \eta' = \cN \,,
\end{equation}
where recall $\eta'$ is the line in $\cT$ to which the Wilson line $W$ is mapped under $D$. This can be seen from the definition of $\cN$ in Figure~\ref{fig: half-space setup}, as follows. Consider fusing this configuration with $\eta'$ from the right-hand side. The discussion of section~\ref{subsec:properties_defects_TYness} implies that moving $\eta'$ past the $D$ interface gives a Wilson line in $\cT/\cG$, which can be absorbed by the $\bra{\text{Dir}}$ boundary condition.

The construction of the self-duality defects does not immediately imply the fusion of $\cN$ with itself, $\cN\times \cN$, which is subtly different from the final fusion rule in \eqref{eq:almostTY}. The fusion rules \eqref{eq:almostTY} and \eqref{eq: Nbar absorb eta} form a subset of those defining the Tambara--Yamagami fusion algebra $TY(\cG)$ \cite{Tambara:1998vmj}, where in addition to \eqref{eq:almostTY} one would also have $\overline{\cN}=\cN$, implying
\begin{align}\label{eq:TY_fusion}
    \cN \times \eta = \cN\qc \cN\times \cN = \sum_{n=1}^k \eta^k\,.
\end{align}
For the line $\cN$ corresponding to the reference isomorphism \eqref{eq: reference iso}, the first of these follows from the fusion rule \eqref{eq:eta' absorb} and the fact that $\cG'=\cG$ for the isomorphism $D$. As we will discuss briefly below, this is a result of our choice of the reference isomorphism \eqref{eq: reference iso}, and fails more generally.

Let us consider the second fusion rule in \eqref{eq:TY_fusion}. For our line $\cN$, scattering twice leaves $\cG$ singlets \eqref{eq:neutrals} invariant, as is seen immediately from \eqref{eq:neutral_scat}. We must therefore have $\cN\times \cN$ proportional to $\sum_{n=1}^k \eta^n$. The only choice of normalisation consistent with the other fusion rules is as given in \eqref{eq:TY_fusion}.\footnote{A proof is as follows. Suppose $\cN\times\cN = c\sum_{n=1}^k \eta^n$ for some constant $c$. We must have $c\in\Z_{\geq0}$ to give sensible fusion rules (indeed, fusion coefficients encode the dimensions of vector spaces associated with junctions between lines \cite{Chang:2018iay}). We now evaluate $\cN\times\cN\times\overline{\cN}$ in two ways. If we fuse $\cN\times\overline{\cN}$ first using \eqref{eq:almostTY} and the first of \eqref{eq:TY_fusion}, we find $\cN\times\cN\times\overline{\cN} = k\cN$. On the other hand, if we fuse $\cN\times\cN$ first and use $\eta\times\overline{\cN}=\overline{\cN}$ (the orientation reversal of the first of \eqref{eq:TY_fusion}) we find $\cN\times\cN\times\overline{\cN} = kc\overline{\cN}$, giving $\cN=c\overline{\cN}$. Similar arguments applied to $\cN\times\overline{\cN}\times\overline{\cN}$ yield $c\cN=\overline{\cN}$, so we conclude that $c=1$ and $\cN$ satisfies \eqref{eq:TY_fusion}.} We conclude, therefore, that the line $\cN$ does have $TY(\cG)$ fusion.

In the present work we will not give a full description of the category of lines in $\cT$. We briefly note, however, that the lines $(1,\eta,\cN)$ must form a $TY(\cG)$ fusion category, the categorical data of which is determined by two further pieces of data known as the \emph{symmetric bicharacter} and the \emph{Frobenius--Schur indicator}. The former is in fact encoded by the value of $r$ found in the calculation of section~\ref{subsec:properties_defects_TYness}, which we elaborate on slightly in Appendix~\ref{app: bicharacter}.

\subsubsection*{Fusion with invertible lines}

Thus far we have chosen a particular isomorphism $D$, given in \eqref{eq: reference iso}, with which to establish the duality $\cT/\cG \to \cT$ and we have constructed the corresponding self-duality defect $\cN$. Other choices of isomorphism $D'$ are related to $D$ by an element of $SO(4)$ as in \eqref{eq:D' from D}, and the corresponding self-duality defects $\cN'$ are related to $\cN$ by \eqref{eq:N' from N}. It follows immediately that the fusion rules \eqref{eq:almostTY} are also satisfied by $\cN'$.

On the other hand, the derivation above that $\cG'=\cG$ for the isomorphism \eqref{eq: reference iso} may not go through for other isomorphisms since the group $\cG'=D(\widehat{\cG})$ depends on $D$. Indeed, one can verify that this relation is not satisfied by the self-duality defects corresponding to the isomorphism \eqref{eq:action_C} for the third and fourth choices of $C$ in \eqref{eq:four_Cs}. In these cases one finds instead that $\cG'=\cG_{(q,p)}\neq\cG_{(p,q)}$.

Relatedly, the $TY(\cG)$ fusion rules \eqref{eq:TY_fusion} may not be satisfied by the other $\cN'$. 
Indeed, the first equation in \eqref{eq:TY_fusion} is only satisfied if $\cG'=\cG$ which, as we have just seen, is not satisfied by the isomorphisms corresponding to the final two choices of $C$ in \eqref{eq:four_Cs}. The second fusion rule in \eqref{eq:TY_fusion} may also not be satisfied for generic isomorphisms $D$. Indeed, it is not satisfied by the final two of \eqref{eq:four_Cs}. The quick way to see this is to note that for these two cases, the matrices $K$ in \eqref{eq:four K's} have $K^2 \neq 1$, which in turn implies that $\cN' \times \cN'$ does not commute with $U(1)^2$, while $\sum_n \eta^n$ clearly does.

Given that the defect $\cN$ corresponding to the isomorphism \eqref{eq:action_C} does satisfy \eqref{eq:TY_fusion}, it is interesting to ask which other $\cN'$ satisfy these fusion rules.
Recall that the action of $\cN$ on $\cG$-neutral operators is determined by its action \eqref{eq:neutral_scat} on the $\cG$-singlets $\Psi_1,\Psi_2$ introduced in \eqref{eq:neutrals}. It is clear that stacking an invertible line $U$ which alters this action to
\begin{equation}
    \Psi_1 \to e^{i \theta} \Psi_2 \qc \Psi_2 \to  e^{-i \theta} \Psi_1\,,
    \label{eq: TY rephasing}
\end{equation}
will still satisfy \eqref{eq:TY_fusion} as performing this map twice still yields a projection onto these $\cG$-neutral operators.

We stress that there is no canonical choice of isomorphism $D:\cT/\cG \to \cT$. While the reference choice \eqref{eq: reference iso} that we made previously satisfies $\cG'=\cG$, it would be equally natural to construct self-duality defects using an isomorphism without this property. The resulting non-invertible lines will always satisfy \eqref{eq:almostTY}, but may not satisfy \eqref{eq:TY_fusion}. We postpone a more detailed discussion of the categorical structure of these defects to future work.

\section{Symmetric boundary conditions}
\label{sec: Symmetric boundary conditions}

In this section we consider conformal boundary conditions for two Dirac fermions which preserve some $U(1)^2$ global symmetry. Such a $U(1)^2$ symmetry must be anomaly-free. We first briefly review the classification of all such symmetries, which fall into two SPT classes, classified by $(I_\Z \Omega^\text{Spin})^3(\text{pt}) \cong \Z_2$. We then show that all such boundary conditions can be constructed using the non-invertible lines $\cN_{(p,q)}$ by folding. Indeed, one needs one of two fundamentally distinct constructions, depending on the SPT class of the $U(1)^2$ to be preserved.

\subsection{Classification of symmetric boundaries}

Let us recall the unique parameterisation of all anomaly-free $U(1)^2$ symmetries of two Dirac fermions \cite{BoyleSmith:2019jnh}. We consider left-movers $\tilde{\psi}_1,\tilde{\psi}_2$ and right-movers $\psi_1,\psi_2$, with charge matrices $\tilde{Q}_{iA}$ and $Q_{iA}$ respectively, with $i = 1,2$ labelling the $U(1)$ symmetry and $A = 1,2$ labelling the species of fermion. Anomalies vanish if
\begin{align}
  \tilde{Q} \tilde{Q}^T -  Q Q^T = 0 \,,
  \label{eq: anomaly-free Q}
\end{align}
which we assume. We also assume $U(1)^2$ is faithful, so $Q$ and $\tilde{Q}$ are of full rank. There is, of course, a degeneracy in this parameterisation as we are free to choose a different basis for $U(1)^2$. The invariant information is neatly captured by the matrix
\begin{align}
  \mathcal{R} = Q^{-1} \tilde{Q}\,,
\end{align}
which is rational and, by the anomaly-free condition \eqref{eq: anomaly-free Q}, orthogonal. The set $\cM = O(2)_\mathbb{Q}$ of such matrices provides a one-to-one parameterisation of anomaly-free faithful $U(1)^2$ symmetries.

For each $\cR \in \cM$, there is an essentially unique simple, conformal boundary condition preserving the $U(1)^2$ symmetry described by $\cR$ \cite{BoyleSmith:2019jnh}. Such boundary conditions come in two SPT classes: Class $\mathcal{V}$, which are interfaces to the trivial phase, and Class $\mathcal{A}$, which are interfaces to the Arf topological phase.\footnote{It is assumed that we have made a suitable choice of sign of the Pauli--Villars regulator mass of the fermions so that the simple linear boundary condition $\psi_i = \tilde{\psi}_i$, preserving the vector-like $U(1)^2$ symmetry, belongs to Class $\cV$. Under a change of sign of the regulator mass of one of the fermions, the theory acquires an Arf, so the classes $\cV$ and $\cA$ are exchanged. In general, if we choose a given quadratic mass term for the Pauli--Villars regulator fields, then the simple linear boundary condition corresponding to that mass term will be in Class $\cV$.} The set $\cM$ of $U(1)^2$ symmetries correspondingly splits into two classes $\cM = \cM_{\mathcal{V}}\cup \cM_{\mathcal{A}}$. The matrices in each class take one of two forms,
\begin{align}
  \cM_{\cV} &= \left\{ \frac{1}{k}\begin{pmatrix}
  	b & -a \\ -a & -b
  \end{pmatrix}\right\} \cup  \left\{     \frac{1}{k}\begin{pmatrix}
  	-a & -b \\ b & -a
  \end{pmatrix} \right\}\,,		\label{eq: MV}\\
    \cM_{\cA} &= \left\{ \frac{1}{k}\begin{pmatrix}
  	-a & -b \\ -b & a
  \end{pmatrix} \right\} \cup \left\{     \frac{1}{k}\begin{pmatrix}
  	-b & a \\ -a & -b
  \end{pmatrix} \right\}\,, \label{eq: MA}
\end{align}
where, in all of these expressions, $a,b,k$ run over the space of coprime integers with $ k>0$, $a^2 + b^2 =k^2$ and $b$ even. Note that in each class the first form for $\cR$ is a reflection, $\cR^2=1$, while the second is a rotation, $\cR^2\neq 1$. We will call a matrix of the first (second) form in \eqref{eq: MV} a reflection (rotation) boundary of Class $\cV$, and similarly for Class $\cA$.

Given a reflection boundary of Class $\cV$, we can immediately get a rotation boundary of Class $\cV$ by stacking with a suitable line in $SO(4)_L\times SO(4)_R$. One such suitable choice is the line $S$ that enacts $(\psi_1,\psi_2) \to (\psi_2,\psi_1)$, since
\begin{align}
  \begin{pmatrix}
  	0 & 1 \\ 1 & 0
  \end{pmatrix}\begin{pmatrix}
  	b & -a \\ -a & -b
  \end{pmatrix} = \begin{pmatrix}
  	-a & -b \\ b & -a
  \end{pmatrix} \,.
\end{align}
The same is true for Class $\cA$, with this same line $S$ in $SO(4)_R$ doing the trick.

We cannot, however, perform a similar transformation between the two classes. Naïvely, given a boundary of Class $\cV$ we might try to get one of Class $\cA$ by stacking the boundary with a line enacting $(\psi_1,\psi_2)\to (\psi_1^\dag,\psi_2)$. But this transformation, viewed as acting on four Majorana--Weyl fermions, lives in $O(4)$ but not $SO(4)$ and thus is not an invertible topological line of the theory (see discussion at the end of section~\ref{subsec: self-duality}). Thus, we have to treat the two classes separately.

\subsection{Boundaries of Class $\cV$}

We can begin with the theory $\cT$ of two right-moving Weyl fermions, and place at $x=0$ the non-invertible defect $\cN_{(p,q)}$. As we have seen, scattering through this interface preserves a $U(1)^2$ acting as in \eqref{eq: non-local U(1) action}.

We then `fold' the theory along the line $x=0$, defining a theory of two Dirac fermions in the half-space $x>0$ where the left-movers are $\tilde{\psi}_i(x) = \psi_i(-x)$. One lands directly on a conformal boundary that preserves a $U(1)^2$ symmetry with charge matrices
\begin{align}
    \tilde{Q} = \begin{pmatrix}
        p & q \\ q & -p
    \end{pmatrix},\qquad Q = \begin{pmatrix}
         q & -p \\ p & q 
    \end{pmatrix} \quad \implies \quad\cR = \frac{1}{k} \begin{pmatrix}
        b & -a \\ -a & -b
    \end{pmatrix} \,.
\label{eq: charge matrices reflection V}
\end{align}
We thus succeed in a generic reflection boundary of Class $\cV$. A left-moving excitation $\tilde{\psi}_1^{m_1}\tilde{\psi}_2^{m_2}$ is scattered into the excitation $\eta^\alpha \psi_1^{n_1}\psi_2^{n_2}$ where $\alpha,n_1,n_2$ are given in \eqref{eq: reference n and alpha}, and the other end of $\eta^{\alpha}$ is anchored on the boundary.

There is an equivalent but sometimes useful way to think about this construction. We can instead start with two Dirac fermions on the half-space $x>0$, with a boundary condition $\bra{V}$ at $x=0$ which imposes the vector-like boundary condition $\tilde{\psi}_i = \psi_i$. We then insert the defect $\cN_{(p,q)}$ at $x=a>0$.  This is depicted below in Figure~\ref{fig: reflection V}. Recall this defect arises from a self-dual gauging of the right-movers $\psi_i$, and thus is transparent to the left-movers.
\begin{center}
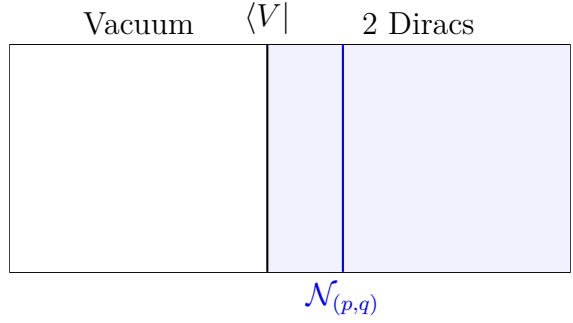
\vspace{1em}
\begin{minipage}{0.8\textwidth}
    \[
        \begin{tikzpicture}
        \draw[thick] (0.6,0) rectangle (8,-3);
        \fill[white] (0.6,0) rectangle (4,-3);
        \fill[blue!5] (4,0) rectangle (8,-3);
        \draw[thick] (4,0) -- (4,-3);
        \draw[thick,blue] (5,0) -- (5,-3);
        \node[above] at (2.3,0) {$\text{Vacuum}$};
        \node[above] at (4,0) {$\bra{V}$};
        \node[above] at (6,0) {2 \text{Diracs}};
        \node[below,blue] at (5,-3) {$\cN_{(p,q)}$};
        \end{tikzpicture}\vspace{-1em}
    \]
    \captionof{figure}{The construction of a reflection boundary of Class $\cV$. We use $\bra{V}$ to denote the vector boundary $\psi_i = \tilde{\psi}_i$. The boundary condition is found by stacking $\cN_{(p,q)}$ onto this vector boundary.} \label{fig: reflection V}
\end{minipage}\vspace{1em}
\end{center}
Scattering off the boundary proceeds in three steps. The left-moving excitation $\tilde{\psi}_1^{m_1} \tilde{\psi}_2^{m_2}$ first passes unscathed through $\cN$, which is completely invisible to the left-moving spectrum. Next, the vector boundary at $x=0$ scatters the excitation into the right-mover $\psi_1^{m_1} \psi_2^{m_2}$. Finally, this right-mover is attached by the topological line $\eta^\alpha$ as it passes back through $\cN$, with the final outgoing state given once again by $\eta^\alpha \psi_1^{n_1}\psi_2^{n_2}$ with $\alpha,n_1,n_2$ as in \eqref{eq: reference n and alpha}. This is summarised in Figure~\ref{fig: reflection V scattering}.

\begin{center}
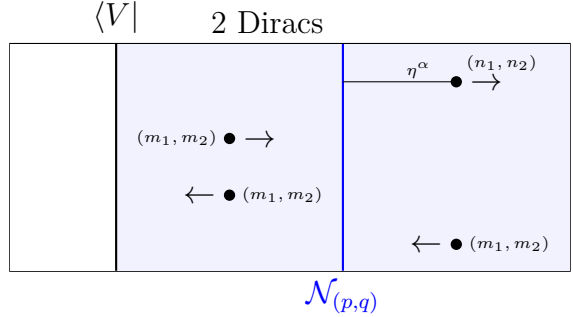
\vspace{1em}
\begin{minipage}{0.8\textwidth}
    \[
        \begin{tikzpicture}
        \draw[thick] (0.6,0) rectangle (8,-3);
        \fill[white] (0.6,0) rectangle (4,-3);
        \fill[blue!5] (2,0) rectangle (8,-3);
        \draw[thick] (2,0) -- (2,-3);
        \draw[thick,blue] (5,0) -- (5,-3);
        \node[above] at (2,0) {$\bra{V}$};
        \node[above] at (4,0) {2 \text{Diracs}};
        \node[below,blue] at (5,-3) {$\cN_{(p,q)}$};
        \node at (6.5,-0.5-2.15) [circle,fill,inner sep=1.5pt]{};
        \node at (6.1,-0.53-2.15) {$\leftarrow$};
        \node at (7.2,-0.5-2.15) {\tiny $(m_1,m_2)$};
        \node at (3.5,-1.25-0.75) [circle,fill,inner sep=1.5pt]{};
        \node at (3.1,-1.28-0.75){$\leftarrow$};
        \node at (4.2,-1.25-0.75) {\tiny $(m_1,m_2)$};
        \node at (3.5,-2+0.75) [circle,fill,inner sep=1.5pt]{};
        \node at (3.9,-2.03+0.75){$\rightarrow$};
        \node at (2.8,-2+0.75) {\tiny $(m_1,m_2)$};
        \node at (6.5,-2.65+2.15) [circle,fill,inner sep=1.5pt]{};
        \node at (6.9,-2.68+2.15) {$\rightarrow$};
        \draw (5,-2.65+2.15)--(6.5,-2.65+2.15);
        \node at (6,-2.5+2.15){\tiny $\eta^\alpha$};
        \node at (7.1,-2.4+2.15){\tiny $(n_1,n_2)$};
        \end{tikzpicture}\vspace{-1em}
    \]
    \captionof{figure}{Scattering off the boundary can be understood in 3 steps, with only the final step non-trivial.}\label{fig: reflection V scattering}
\end{minipage}\vspace{1em}
\end{center}
At this point we can take $a\to 0$, by the topological nature of $\cN$. We can thus realise the simple conformal boundary preserving \eqref{eq: charge matrices reflection V} as a straightforward vector boundary dressed with the self-duality defect $\cN_{(p,q)}$.

Following the discussion above, we can obtain a generic rotation boundary of Class $\cV$ by further dressing this reflection boundary with the invertible line $S$, which we recall is the $SO(4)$ line enacting the swap $(\psi_1,\psi_2)\to (\psi_2,\psi_1)$. The resulting boundary preserves a $U(1)^2$ symmetry
\begin{align}
    \tilde{Q} = \begin{pmatrix}
        p & q \\ q & -p
    \end{pmatrix},\qquad Q = \begin{pmatrix}
         -p & q \\q & p  
    \end{pmatrix} \quad \implies \quad\cR = \frac{1}{k} \begin{pmatrix}
        -a & -b \\ b & -a
    \end{pmatrix} \,.
\end{align}
This construction is then depicted below in Figure~\ref{fig: rotation V}.
\begin{center}\vspace{1em}
\begin{minipage}{0.8\textwidth}
    \[
        \begin{tikzpicture}
        \draw[thick] (0.6,0) rectangle (8,-3);
        \fill[white] (0.6,0) rectangle (4,-3);
        \fill[blue!5] (4,0) rectangle (8,-3);
        \draw[thick] (4,0) -- (4,-3);
        \draw[thick,blue] (5,0) -- (5,-3);
        \draw[thick,orange] (6,0) -- (6,-3);
        \node[above] at (2.3,0) {$\text{Vacuum}$};
        \node[above] at (4,0) {$\bra{V}$};
        \node[above] at (6,0) {2 \text{Diracs}};
        \node[below,blue] at (5,-3) {$\cN_{(p,q)}$};
        \node[below,orange] at (6,-3) {$S$};
        \end{tikzpicture}\vspace{-1em}
    \]
    \captionof{figure}{The construction of a rotation boundary of Class $\cV$.} \label{fig: rotation V}
\end{minipage}\vspace{1em}
\end{center}
We have thus succeeded in constructing all boundaries of Class $\cV$.

\subsection{Boundaries of Class $\cA$}\label{subsec: Class A}

We need an additional ingredient to construct boundaries of Class $\cA$. Suppose we start with a vector boundary stacked with $\cN_{(p,q)}$. What we would like to do is further stack this boundary with a line $S'$ that generates $(\psi_1,\psi_2)\to (\psi_2^\dagger,\psi_1)$. The resulting boundary preserves the $U(1)^2$ symmetry
\begin{align}
    \tilde{Q} = \begin{pmatrix}
        p & q \\ q & -p
    \end{pmatrix},\qquad Q = \begin{pmatrix}
        -p&-q  \\ q&-p  
    \end{pmatrix} \quad \implies \quad\cR = \frac{1}{k} \begin{pmatrix}
        -a&-b  \\ -b&a 
    \end{pmatrix}\,,
\label{eq: preserved symm reflection A}
\end{align}
and thus is a reflection boundary of Class $\cA$. The catch is that $S'$, when viewed as acting on four right-moving Majorana--Weyl fermions, belongs to $O(4)$ but not $SO(4)$. As such, it is not an invertible line of the theory $\cT$ of two right-moving Weyl fermions. Instead, it is an isomorphism (or invertible interface)
\begin{align}
  \cT \otimes \text{Arf} \xrightarrow{\,\,\,S'\,\,\,} T \,,
\end{align}
as shown in \eqref{eq:Arf_absorb}.

Thus attempting to stack with the line $S'$ produces the setup depicted below in Figure~\ref{fig: reflection A}, which does define a boundary condition preserving the symmetry \eqref{eq: preserved symm reflection A}, but of the wrong theory.
\begin{center}
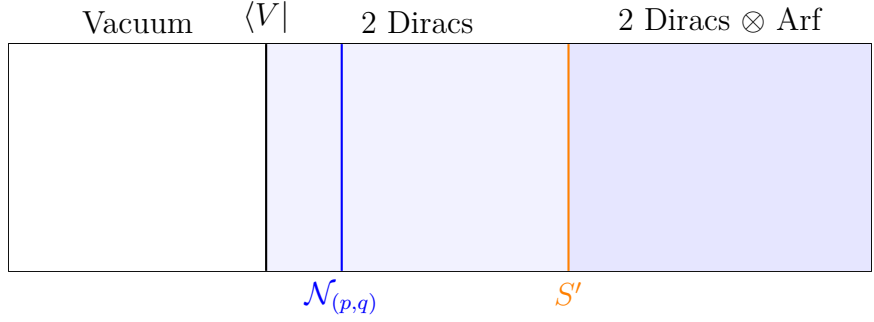
\vspace{1em}
\begin{minipage}{\textwidth}
    \[
        \begin{tikzpicture}
        \draw[thick] (0.6,0) rectangle (12,-3);
        \fill[white] (0.6,0) rectangle (4,-3);
        \fill[blue!5] (4,0) rectangle (8,-3);
        \fill[blue!10] (8,0) rectangle (12,-3);
        \draw[thick] (4,0) -- (4,-3);
        \draw[thick,blue] (5,0) -- (5,-3);
        \draw[thick,orange] (8,0) -- (8,-3);
        \node[above] at (2.3,0) {Vacuum};
        \node[above] at (4,0) {$\bra{V}$};
        \node[above] at (6,0) {$2$ Diracs};
        \node[above] at (10,0) {$2$ Diracs $\otimes$ Arf };
        \node[below,blue] at (5,-3) {$\cN_{(p,q)}$};
        \node[below,orange] at (8,-3) {$S'$};
        \end{tikzpicture}\vspace{-1em}
    \]
    \captionof{figure}{The effect of stacking with the interface $S'$.}\label{fig: reflection A}
\end{minipage}\vspace{1em}
\end{center}
To rectify this, we can make use of the fact that Arf is an invertible phase, with $\text{Arf} \otimes \text{Arf} = 1$. Therefore stacking the entire Figure~\ref{fig: reflection A} with the Arf phase produces the setup shown below in Figure~\ref{fig: reflection A2}.
\begin{center}
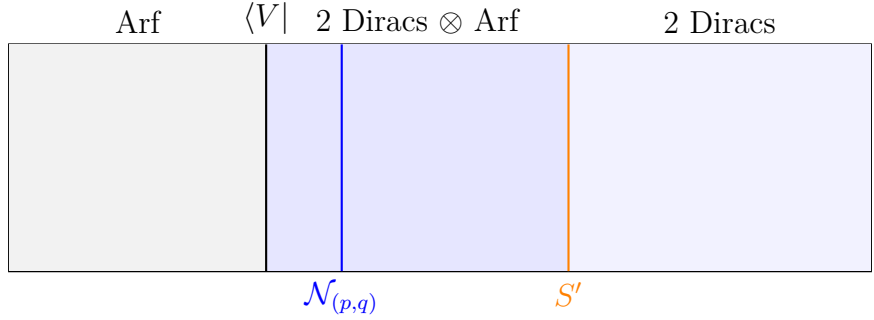
\vspace{1em}
\begin{minipage}{\textwidth}\centering
\begin{minipage}{\textwidth}
    \[
        \begin{tikzpicture}
        \draw[thick] (0.6,0) rectangle (12,-3);
        \fill[black!5] (0.6,0) rectangle (4,-3);
        \fill[blue!10] (4,0) rectangle (8,-3);
        \fill[blue!5] (8,0) rectangle (12,-3);
        \draw[thick] (4,0) -- (4,-3);
        \draw[thick,blue] (5,0) -- (5,-3);
        \draw[thick,orange] (8,0) -- (8,-3);
        \node[above] at (2.3,0) {Arf};
        \node[above] at (4,0) {$\bra{V}$};
        \node[above] at (6,0) {$2$ Diracs $\otimes$ Arf};
        \node[above] at (10,0) {$2$ Diracs};
        \node[below,blue] at (5,-3) {$\cN_{(p,q)}$};
        \node[below,orange] at (8,-3) {$S'$};
        \end{tikzpicture}\vspace{-1em}
    \]
    \end{minipage}\vspace{1em}
    \begin{minipage}{0.8\textwidth}
    \captionof{figure}{The construction of a reflection boundary of Class $\cA$. This is an interface to the Arf theory, rather than a true boundary condition.}\label{fig: reflection A2}
\end{minipage}
\end{minipage}
\end{center}
As usual, one finds the desired boundary condition by stacking all lines on top of $\bra{V}$. The resulting configuration describes an interface from two Dirac fermions to Arf, rather than to the vacuum, so is not a boundary condition in the formal sense.

For completeness, we finally need to construct a boundary condition of the fourth and final type: a rotation boundary of Class $\cA$. This is easy: we just need to dress a reflection boundary of Class $\cA$ with the invertible line $S$, as depicted below in Figure~\ref{fig: rotation A}.
\begin{center}\vspace{1em}
\begin{minipage}{\textwidth}\centering
\begin{minipage}{\textwidth}
    \[
        \begin{tikzpicture}
        \draw[thick] (0.6,0) rectangle (12,-3);
        \fill[black!5] (0.6,0) rectangle (4,-3);
        \fill[blue!10] (4,0) rectangle (8,-3);
        \fill[blue!5] (8,0) rectangle (12,-3);
        \draw[thick] (4,0) -- (4,-3);
        \draw[thick,blue] (5,0) -- (5,-3);
        \draw[thick,orange] (8,0) -- (8,-3);
        \draw[thick,orange] (9,0) -- (9,-3);
        \node[above] at (2.3,0) {Arf};
        \node[above] at (4,0) {$\bra{V}$};
        \node[above] at (6,0) {$2$ Diracs $\otimes$ Arf};
        \node[above] at (10,0) {$2$ Diracs};
        \node[below,blue] at (5,-3) {$\cN_{(p,q)}$};
        \node[below,orange] at (8,-3) {$S'$};
        \node[below,orange] at (9,-3) {$S$};
        \end{tikzpicture}\vspace{-1em}
    \]
    \end{minipage}\vspace{1em}
    \begin{minipage}{0.8\textwidth}
    \captionof{figure}{The construction of a rotation boundary of Class $\cA$.} \label{fig: rotation A}
\end{minipage}
\end{minipage}\vspace{-0.5em}
\end{center}
The resulting boundary preserves the $U(1)^2$ symmetry
\begin{align}
    \tilde{Q} = \begin{pmatrix}
        p & q \\ q & -p
    \end{pmatrix},\qquad Q = \begin{pmatrix}
         -q & -p \\ -p & q  
    \end{pmatrix} \quad \implies \quad\cR = \frac{1}{k} \begin{pmatrix}
        -b & a \\ -a & -b
    \end{pmatrix}\,,
\end{align}
which is indeed the $\cR$ matrix of a rotation boundary of Class $\cA$. As above, however, this configuration describes an interface to the Arf theory, rather than the vacuum.

\subsubsection*{An alternative perspective: lines dressed with Majoranas}

We have found that we can build `boundary conditions' of Class $\cA$ as in Figures~\ref{fig: reflection A2} and \ref{fig: rotation A}, but they are interfaces not to the vacuum but to the Arf topological phase. They are therefore not technically boundary conditions.

There is an alternative way we can think about this: we \emph{can} build an interface from two Dirac fermions to the vacuum, a genuine boundary condition of the theory, but this necessitates the inclusion on the boundary of a single unpaired Majorana mode. This is because the Arf theory admits a topological boundary which hosts such a Majorana mode \cite{Kitaev:2000nmw}, which allows us to glue Figure~\ref{fig: reflection A2} to the following setup:
\begin{center}
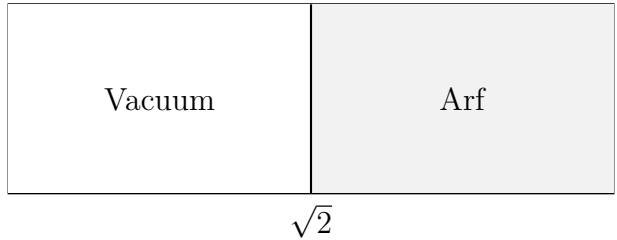
\vspace{1em}
\begin{minipage}{0.8\textwidth}
    \[
        \begin{tikzpicture}
        \draw[thick] (0,0) rectangle (8,-2.5);
        \fill[white] (0,0) rectangle (4,-2.5);
        \fill[black!5] (4,0) rectangle (8,-2.5);
        \draw[thick] (4,0) -- (4,-2.5);
        \node at (2,-1.25) {Vacuum};
        \node[below] at (4,-2.5) {$\sqrt{2}$};
        \node at (6,-1.25) {Arf};
        \end{tikzpicture}\vspace{-1em}
    \]
    \captionof{figure}{The Arf topological phase with an unpaired Majorana fermion on its boundary, which is a well-defined two-dimensional system without anomaly. We denote the unpaired Majorana mode by $\sqrt{2}$, for the factor of $\sqrt{2}$ it contributes to the partition function.} \label{fig: arf boundary}
\end{minipage}\vspace{1em}
\end{center}
For a reflection of Class $\cA$ the relevant picture is shown below in Figure~\ref{fig: reflection A sqrt2}:
\begin{center}
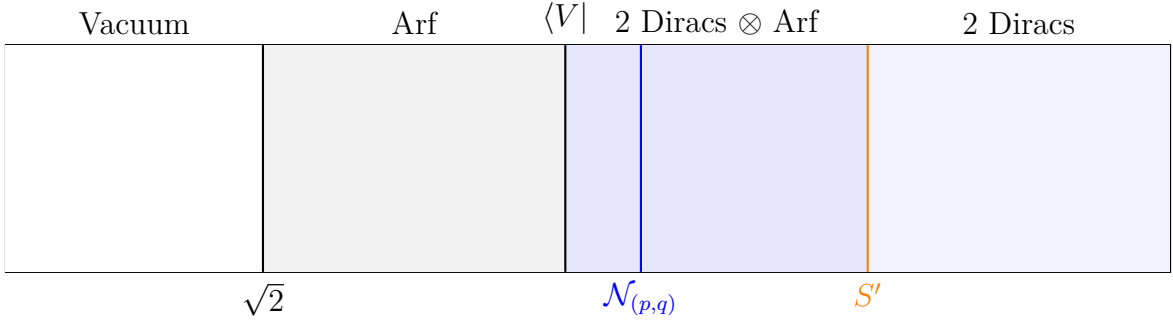
\vspace{1em}
\begin{minipage}{\textwidth}\centering
\begin{minipage}{\textwidth}
    \[
        \begin{tikzpicture}
        \draw[thick] (-3.4,0) rectangle (12,-3);
        \fill[white] (-3.4,0) rectangle (0,-3);
        \fill[black!5] (0,0) rectangle (4,-3);
        \fill[blue!10] (4,0) rectangle (8,-3);
        \fill[blue!5] (8,0) rectangle (12,-3);
        \draw[thick] (0,0) -- (0,-3);
        \draw[thick] (4,0) -- (4,-3);
        \draw[thick,blue] (5,0) -- (5,-3);
        \draw[thick,orange] (8,0) -- (8,-3);
        \node[above] at (-1.7,0) {Vacuum};
        \node[above] at (2,0) {Arf};
        \node[above] at (4,0) {$\bra{V}$};
        \node[above] at (6,0) {$2$ Diracs $\otimes$ Arf};
        \node[above] at (10,0) {$2$ Diracs};
        \node[below] at (0,-3) {$\sqrt{2}$};
        \node[below,blue] at (5,-3) {$\cN_{(p,q)}$};
        \node[below,orange] at (8,-3) {$S'$};
        \end{tikzpicture}\vspace{-1em}
    \]
    \end{minipage}\vspace{1em}
    \begin{minipage}{0.8\textwidth}
    \captionof{figure}{The construction of a non-simple reflection boundary of Class $\cA$ involves an unpaired Majorana mode on the boundary.} \label{fig: reflection A sqrt2}
\end{minipage}
\end{minipage}
\end{center}
Upon collapsing all interfaces together, we obtain a true conformal boundary of two Dirac fermions preserving the Class $\cA$ symmetry \eqref{eq: preserved symm reflection A}. However, the price to pay is that the boundary is no longer simple: the boundary topological operators comprise not just the identity operator, but also insertions of the unpaired Majorana mode $\psi$.

In the language of super-fusion categories, boundaries such as Figure~\ref{fig: reflection A sqrt2} which host one bosonic and one fermionic topological operator are viewed as simple in a generalised sense, and called \emph{q-type} \cite{Chang_2023}. The more familiar simple boundaries such as Figure~\ref{fig: reflection V}, hosting only a single bosonic topological operator, are known as \emph{m-type}.

A similar picture applies for rotations of Class $\cA$, which feature an additional line $S$ on the right.

\section{Microscopic descriptions}\label{sec: UV}

In this final section, we explore some microscopic setups which flow at low energies to the defects $\cN_{(p,q)}$ and their associated boundary conditions.

\subsection{Non-invertible lines from rotors}

The simplest of these comes in the form of fermions coupled to a quantum-mechanical rotor degree of freedom \cite{Polchinski:1984uw}. Indeed, the precise rotor model which flows to the line $\cN_{(p,q)}$ corresponding to the $\Z_k$ gauging \eqref{eq: Zk action} was studied recently in \cite{Loladze:2025jsq}. Let us review this now.

One considers two Weyl fermions coupled to a periodic scalar degree of freedom $\sigma \sim \sigma+2\pi$ that is localised to $x=0$, with charges $v = (v_1, v_2) = (p-q, p+q)$. The action is
\begin{align}
  S = \int d^2 x \left(i \psi_1^\dagger \partial_+ \psi_1 + i \psi_2^\dagger \partial_+ \psi_2 \right) +\int dt \left(\frac{I}{2}\dot{\sigma}^2 +\sigma\!\left(v_1\psi_1^\dagger \psi_1 + v_2\psi_2^\dagger \psi_2\right)\right) \,.
  \label{eq: rotor}
\end{align}
An application of \cite{Polchinski:1984uw, Loladze:2025jsq} then shows that the IR limit of the rotor is a topological line defect, and that it preserves a chiral $U(1)^2$ acting differently on the fermions to the left and the right of the rotor with charge matrices
\begin{align}
    \tilde{Q} = \begin{pmatrix}
        v_1 & v_2 \\ v_2 & -v_1
    \end{pmatrix},\quad Q = \begin{pmatrix}
        -v_1 & -v_2 \\ v_2 & -v_1
    \end{pmatrix} \; \implies \; \cR = \mathds{1} - \frac{2v v^T}{v^2} \,.
\label{eq: charge matrices rotor}
\end{align}
For this construction to work, it is particularly important that the charge vector $v$ has even length-squared, which it does by virtue of $v^2 = 2k$, otherwise the rotor would suffer from a fermion parity anomaly and require the attachment of a bulk Arf phase, giving an interface $\cT \to \cT \otimes \text{Arf}$ instead.\footnote{This can be seen by applying the diagnostic of \cite{BoyleSmith:2019jnh} to compute the SPT class of the boundary state associated to \eqref{eq: charge matrices rotor}, which yields class $\cV$ or class $\cA$ depending on whether $v^2$ is even or odd.} This is why we could not choose the charge vector to be $v = (q, -p)$, for example, since exactly one of $p, q$ is odd.

Given then that the rotor flows to a genuine defect of $\cT$, we now invoke the further fact that $v_1$ and $v_2$ are coprime, which is believed to imply it flows to a \emph{simple} defect. This almost forces it to be $\cN_{(p,q)}$. Indeed, the simple boundaries preserving \eqref{eq: charge matrices rotor} were classified by \cite{BoyleSmith:2019jnh}, and after unfolding to interfaces, all are of the form $\cN_{(p,q)}$ times an invertible $U(1)^2$ line, as we've discussed above. As a further confirmation, we can check that the symmetry \eqref{eq: charge matrices rotor} contains both a left-acting and a right-acting chiral $\cG_{(p,q)}$, which $\cN_{(p,q)}$ absorbs. Thus modulo our simplicity assumption, and up to fusion with $U(1)^2$ lines, all $\cN_{(p,q)}$ defects have a UV realisation by an abelian rotor.

We point out that the rotor need not always land on the defect $\cN_{(p,q)}$ directly. For example, in one of the original examples $(p, q) = (1, 0)$ considered by Polchinski \cite{Polchinski:1984uw}, the scattering of fermions through the rotor was computed to be $\psi_i \to e^{i \theta} \epsilon_{ij} \psi_j^\dagger$ where $\epsilon_{ij}$ is antisymmetric with $\epsilon_{12} = 1$, and $\theta$ is a free parameter. This can also be understood as the unique scattering that preserves $SU(2)$ symmetry but flips $U(1)$ charge. Then scattering twice sends $\psi_i \to -\psi_i$, or equivalently two copies of the rotor fused together gives the $(-1)^F$ line, not the identity. Thus the definition of $\cN_{(p,q)}$ we have chosen, which uses the freedom $\cN \to \cN \times U$ to ensure that $\cN$ is Tambara--Yamagami, need not coincide with the rotor.

\subsection{Boundaries from symmetric mass generation}\label{sec:SMG}

We have seen that we can only construct a simple conformal boundary (that is, a topological interface to the trivial theory) preserving symmetries of Class $\cV$. For symmetries of Class $\cA$, we are stuck with either a non-simple boundary, or a simple topological interface not to the trivial theory, but the Arf SPT.

It is natural to ask whether this dichotomy between Classes $\cV$ and $\cA$ also rears its head when we consider symmetric mass generation for $\cT_\text{Dir}$, the theory of two Dirac fermions, since doing so in a half-space should dynamically generate corresponding symmetric boundary conditions. The answer, as we now show, is yes. This follows by a generalisation of the analysis of \cite{Mouland:2025ilu}. 

\subsubsection*{SMG for Class $\cV$}

We first consider gapping $\cT_\text{Dir}$ while preserving a symmetry of Class $\cV$. By a suitable $SO(4)$ transformation we can take without loss of generality the charges
\begin{align}
    \begin{array}{c|cc:cc}
        & \tilde{\psi}_{1} & \tilde{\psi}_{2} & \psi_1 & \psi_2 \\ \hline
        U(1)_1 & p & q  & q & -p\\
        U(1)_2 & q & -p & p & q \\
    \end{array}\qquad\implies \quad\cR = \frac{1}{k} \begin{pmatrix}
        b & -a \\ -a & -b
    \end{pmatrix}\,.
\label{eq: SMG charges}
\end{align}
A procedure to gap $\cT_\text{Dir}$ while preserving this symmetry was studied in \cite{Tong:2021phe, Mouland:2025ilu}. One first writes down a different theory $\widehat{\cT}_\text{Dir}$, which is found from $\cT_\text{Dir}$ by gauging $U(1)_1\times U(1)_2$ and adding two complex scalar fields $\phi_1,\phi_2$. Then $\widehat{T}$ has a gauge symmetry $\widehat{U(1)}_1\times \widehat{U(1)}_2$ along with a global symmetry $U(1)_1\times U(1)_2$, under which the fields have the following charges:
\begin{align}
    \begin{array}{c|cc:cc:cc}
        & \tilde{\psi}_{1} & \tilde{\psi}_{2} & \psi_1 & \psi_2 & \phi_1 & \phi_2  \\\hline
       \rule{0pt}{3ex} \widehat{U(1)}_1 & p & q  & q & -p & 1 & 0		\\
        \widehat{U(1)}_2 & q & -p & p & q & 0 & 1		\\
        U(1)_1 & 0 & 0 & 0 & 0 & -1 & 0					\\
        U(1)_2 & 0 & 0 & 0 & 0 & 0 & -1
    \end{array}
\end{align}
The idea then is to study the low energy physics of $\widehat{\cT}_\text{Dir}$ as we vary the masses $m_1^2,m_2^2$ of the scalar fields. In what follows we denote by $e$ the bare UV gauge coupling, which for simplicity we can take to be equal for the two gauge fields.

The Higgs phase $m_1^2,m_2^2 \ll -e^2$ is straightforward: the scalar fields condense, gapping out the two gauge fields. Meanwhile the global $U(1)_1\times U(1)_2$ survives after being twisted with the gauge symmetry. Crucially the two fermions remain gapless, precisely because the chiral nature of the gauge symmetry means that there are no gauge-invariant Yukawa couplings that can be dynamically generated. We land precisely on the theory $\cT_\text{Dir}$ with the charge assignments \eqref{eq: SMG charges}. The stability of this Higgs phase---that is, its gaplessness down to some finite critical VEVs for the scalar fields---was verified via a bosonisation analysis in \cite{Mouland:2025ilu}.

Less trivial is the confining phase $m_1^2,m_2^2 \gg e^2$, in which the scalar fields are gapped out and we are left with a gauged version of $\cT_\text{Dir}$. The fact that the rank of the gauge group equals the number of fermions ensures on general grounds that this phase is gapped. But the nature of this gapped phase is not immediate. The NS-NS partition function of $\widehat{\cT}_\text{Dir}$ in this phase was determined in \cite{Mouland:2025ilu}. We extend this result here to generic spin structure.
 
The first step is to pass to a bosonised description of $\cT_\text{Dir}$. This consists of a pair of compact bosons $\varphi_1\sim \varphi_1+2\pi$, $\varphi_2 \sim \varphi_2+2\pi$ with action
 \begin{align}
  S =\int  \left( \frac{1}{8\pi}(d \varphi_1+2qA_2)^2+\frac{1}{8\pi}(d \varphi_2-2pA_2)^2- \frac{1}{2\pi}\left(p\varphi_1 + q\varphi_2\right) dA_1 \right) \,,
\label{eq: boson theory}
\end{align}
where here $A_1,A_2$ are background fields for $U(1)_1,U(1)_2$, respectively, which couple as \eqref{eq: SMG charges} in the fermionic description.\footnote{\label{fn:bosonisation}There is a freedom in how to assemble our four Weyl fermions into two Dirac fermions before bosonisation. Using the conventions of \cite[eq.~(3.1)]{Mouland:2025ilu} the couplings in \eqref{eq: boson theory} are found by bosonising the Dirac fermions $(\tilde{\psi}_1,\psi_2^\dag)$ and $(\tilde{\psi}_2,\psi_1)$. The single conjugation involved acts on the four Majorana--Weyl fermions as an element of $O(4)$ which is not in $SO(4)$, introducing the extra factor of $(-1)^{\text{Arf}[\rho]}$ at the front of \eqref{eq: bosonisation} (see the discussion at the end of section~\ref{subsec: self-duality}).} The fermionic theory is then recovered by gauging the $\Z_2$ subgroups of the $U(1)$ winding symmetry of each boson, and coupling the resulting $\Z_2$ gauge fields via the Arf invariant to the choice of spin structure \cite{Karch:2019lnn}.
 
Let's see this more explicitly. Let $Z_{\text{Dir},\cV}^{(u,v)}[A_1,A_2]$ denote the partition function of $\cT_\text{Dir}$ with spin structure $\rho=(u,v)$, with coupling of background fields $A_1,A_2$ dictated by \eqref{eq: SMG charges}. Meanwhile let $Z_\text{bos}^{(\alpha_i,\beta_i)}[A_1,A_2]$, $\alpha_i,\beta_i=0,1$, denote the partition function of the theory \eqref{eq: boson theory} with twists in the winding symmetry\footnote{These twists are enacted by the insertion of defects $\frac{1}{2} \int \frac{1}{2\pi} d\varphi_i $ around the dual cycle. One can equivalently work in a T-dual frame, in which they become twisted boundary conditions.} of $\varphi_i$ by $(-1)^{\alpha_i}$ around the spatial cycle and by $(-1)^{\beta_i}$ around the temporal cycle. One then has precisely
 \begin{align}
  Z_{\text{Dir},\cV}^{(u,v)}[A_1,A_2] = \frac{1}{4}(-1)^{uv}  \sum_{\alpha_1,\beta_1,\alpha_2,\beta_2} (-1)^{(\alpha_1+u)(\beta_1 + v)+(\alpha_2+u)(\beta_2 + v)} Z_\text{bos}^{(\alpha_i,\beta_i)}[A_1,A_2] \,.
\label{eq: bosonisation}
\end{align}
We recognise the prefactor as $(-1)^{\text{Arf}[\rho]}$, while the sign appearing inside the sum is $(-1)^{\text{Arf}[\rho + a_1]+\text{Arf}[\rho + a_2]}$ where $a_i$ is a $\Z_2$ gauge field with holonomies $(\alpha_i,\beta_i)$.\footnote{A spin structure $\rho$ determines the periodicity of fermions around framed 1-cycles. Given a $\Z_2$ gauge field $a$ with holonomies $\oint a = 0,1$ we define the spin structure $\rho + a$ to be the same as $\rho$ on cycles where $\oint a=0$ and opposite to $\rho$ on cycles where $\int a = 1$. That is, the periodicity of fermions is exchanged from periodic to anti-periodic and vice versa on such cycles.}

We also then pass to a bosonised description of the gauge theory $\widehat{\cT}_\text{Dir}$. This is given by
 \begin{align}
  \hat{S} =\int  \bigg(& -\frac{1}{2e^2}(da_1)^2 - \frac{1}{2e^2}(da_2)^2+ \frac{1}{8\pi}(d \varphi_1+2qa_2)^2+\frac{1}{8\pi}(d \varphi_2-2pa_2)^2\nn\\
  &- \frac{1}{2\pi}\left(p\varphi_1 + q\varphi_2\right) da_1 + |d\phi_1  +i(a_1-A_1) \phi_1|^2 + |d\phi_2 +i(a_2-A_2) \phi_2|^2 \nn\\
  &-m_1^2 |\phi_1|^2 - m_2^2 |\phi_2|^2 - \lambda_1|\phi_1|^4 - \lambda_2 |\phi_2|^4\bigg) \, .
\label{eq: boson theory gauged}
\end{align}
Here $a_i$ are dynamical gauge fields for $\widehat{U(1)}_i$, while as before $A_i$ are background gauge fields for $U(1)_i$. Writing $\hat{Z}_\text{bos}^{(\alpha_i,\beta_i)}[A_1,A_2]$ for the partition function of \eqref{eq: boson theory gauged} with the same twists as before in the winding symmetries of $\varphi_i$, and $\hat{Z}_{\text{Dir},\cV}^{(u,v)}[A_1,A_2]$ for the partition function of $\widehat{\cT}_\text{Dir}$, we have then
 \begin{align}
  \hat{Z}_{\text{Dir},\cV}^{(u,v)}[A_1,A_2] = \frac{1}{4}(-1)^{uv}  \sum_{\alpha_1,\beta_1,\alpha_2,\beta_2} (-1)^{(\alpha_1+u)(\beta_1 + v)+(\alpha_2+u)(\beta_2 + v)} \hat{Z}_\text{bos}^{(\alpha_i,\beta_i)}[A_1,A_2] \,.
\label{eq: bosonisation gauged}
\end{align}
We then consider the confining phase $m_1^2,m_2^2\gg e^2$. We can integrate out the heavy $\phi_1,\phi_2$ to land on a $U(1)^2$ gauge theory coupled to two compact bosons, where we see all dependence on $A_i$ has gone. It was then shown in \cite{Mouland:2025ilu} that in this phase, the theory \eqref{eq: boson theory gauged} flows at low energies to 
\begin{align}
  \hat{Z}_\text{bos}^{(\alpha_i,\beta_i)}[A_1,A_2] \to 2\delta_{q\alpha_1 - p\alpha_2,0}\delta_{q\beta_1 - p \beta_2,0}\,,
\end{align}
where these Kronecker symbols are understood modulo $2$. Hence, we find
 \begin{align}
  \hat{Z}_{\text{Dir},\cV}^{(u,v)}[A_1,A_2] &\to  \frac{1}{2}(-1)^{uv}  \sum_{\alpha_1,\beta_1,\alpha_2,\beta_2} (-1)^{(\alpha_1+u)(\beta_1 + v)+(\alpha_2+u)(\beta_2 + v)} \delta_{q\alpha_1 - p\alpha_2,0}\delta_{q\beta_1 - p \beta_2,0}		\nn\\
  &\qquad =(-1)^{uv}(-1)^{uv}		\nn\\
  &\qquad = 1\,.
\end{align}
And thus, in the confining phase, $\widehat{\cT}_\text{Dir}$ flows to the trivial theory.

We propose then that we can dynamically generate the boundary condition described by Figure~\ref{fig: reflection V} by taking the gauge theory $\widehat{\cT}_\text{Dir}$ and spatially modulating the masses $m_1^2,m_2^2$ such that in the region $x<0$ it lies in the confining phase $m_1^2,m_2^2 \gg e^2$, while for $x>0$ it is in the Higgs phase $m_1^2,m_2^2\ll -e^2$.

\subsubsection*{SMG for Class $\cA$}

We are then left to consider trying to gap $\cT_\text{Dir}$ while preserving a symmetry of Class $\cA$. This proceeds very similarly. Without loss of generality we can choose our Class $\cA$ symmetry to act as
\begin{align}
    \begin{array}{c|cc:cc}
        & \tilde{\psi}_{1} & \tilde{\psi}_{2} & \psi_1 & \psi_2 \\ \hline
        U(1)_1 & p & q  & -p & -q\\
        U(1)_2 & q & -p & q & -p \\
    \end{array}\qquad\implies \quad\cR = \frac{1}{k} \begin{pmatrix}
        -a & -b \\ -b & a
    \end{pmatrix} \,.
\label{eq: SMG charges A}
\end{align}
The gauge theory $\widehat{\cT}_\text{Dir}$ that enacts SMG is then given by
\begin{align}
    \begin{array}{c|cc:cc:cc}
        & \tilde{\psi}_{1} & \tilde{\psi}_{2} & \psi_1 & \psi_2 & \phi_1 & \phi_2 \\ \hline
        \rule{0pt}{3ex}\widehat{U(1)}_1 & p & q  & -p & -q & 1 & 0		\\
        \widehat{U(1)}_2 & q & -p & q & -p & 0 & 1		\\
        U(1)_1 & 0 & 0 & 0 & 0 & -1 & 0					\\
        U(1)_2 & 0 & 0 & 0 & 0 & 0 & -1
    \end{array}
\label{eq: SMG A}
\end{align}
We can once again pass to a bosonised description, which we can in fact take as precisely the \textit{same} bosonic theory \eqref{eq: boson theory}.\footnote{Here we are again exploiting the freedom to rearrange our four Weyl fermions into two Dirac fermions in different ways. In particular, the couplings in \eqref{eq: boson theory} are found in this case by bosonising the Dirac fermions $(\tilde{\psi}_1,\psi_1^\dag)$ and $(\tilde{\psi}_2,\psi_2^\dag)$. The extra conjugation required here with respect to the Class $\cV$ case (see footnote~\ref{fn:bosonisation}) is responsible for the relative phase $(-1)^{\text{Arf}[\rho]}$ between \eqref{eq: bosonisation} and \eqref{eq: bosonisation A}.} The analogues of \eqref{eq: bosonisation} and \eqref{eq: bosonisation gauged} are
 \begin{align}
  Z_{\text{Dir},\cA}^{(u,v)}[A_1,A_2] = \frac{1}{4}  \sum_{\alpha_1,\beta_1,\alpha_2,\beta_2} (-1)^{(\alpha_1+u)(\beta_1 + v)+(\alpha_2+u)(\beta_2 + v)} Z_\text{bos}^{(\alpha_i,\beta_i)}[A_1,A_2]\,,
\label{eq: bosonisation A}
\end{align}
and 
 \begin{align}
  \hat{Z}_{\text{Dir},\cA}^{(u,v)}[A_1,A_2] = \frac{1}{4}  \sum_{\alpha_1,\beta_1,\alpha_2,\beta_2} (-1)^{(\alpha_1+u)(\beta_1 + v)+(\alpha_2+u)(\beta_2 + v)} \hat{Z}_\text{bos}^{(\alpha_i,\beta_i)}[A_1,A_2]\,,
\label{eq: bosonisation gauged A}
\end{align}
which just differ by their lack of $(-1)^{\text{Arf}[\rho]}$ prefactor. The Higgs phase works just the same, landing on $\widehat{\cT}_\text{Dir}$ with the Class $\cA$ charge assignments \eqref{eq: SMG A}. In contrast, in the confining phase we find
 \begin{align}
  \hat{Z}_{\text{Dir},\cA}^{(u,v)}[A_1,A_2] &\to  \frac{1}{2}  \sum_{\alpha_1,\beta_1,\alpha_2,\beta_2} (-1)^{(\alpha_1+u)(\beta_1 + v)+(\alpha_2+u)(\beta_2 + v)} \delta_{q\alpha_1 - p\alpha_2,0}\delta_{q\beta_1 - p \beta_2,0}		\nn\\
  &\qquad =(-1)^{uv}		\nn\\
  &\qquad = (-1)^{\text{Arf}[\rho]} \,.
\end{align}
Thus, as predicted, we find that the SMG model \eqref{eq: SMG A} does \textit{not} successfully deform $\cT_\text{Dir}$ to a trivially gapped phase while preserving this Class $\cA$ symmetry, but does succeed in deforming it to the Arf theory.

Once again, by spatially modulating the masses $m_1^2,m_2^2$, we can dynamically generate the interface to the Arf theory in Figure~\ref{fig: reflection A2}.

We have learnt that the SMG model $\widehat{\cT}_\text{Dir}$ in fact fails to trivially gap the theory. This is however easily rectified by the addition of one more ingredient. We can instead consider the UV theory $\widehat{\cT}_\text{Dir}\otimes \text{Maj}$, which is just this gauge theory tensored with a single, decoupled Majorana fermion with mass $m$. In the Higgs phase we take $m\gg e$, such that we still land on the theory $\cT_\text{Dir}$. In contrast, in the confining phase, if we also take $m\ll -e$ then integrating out this Majorana mode generates an additional overall factor of $(-1)^{\text{Arf}[\rho]}$, and thus the theory is now trivially gapped in this phase. Crucially, doing this now in a spatially modulated way, we must pass through $m=0$ and thus the resulting boundary condition is dressed with a single unpaired Majorana mode; this is precisely Figure~\ref{fig: reflection A sqrt2}.

\section{Discussion}\label{sec: discussion}

We have studied the relation of categorical symmetries of two Weyl fermions and symmetric boundary conditions of two Dirac fermions in 1+1 dimensions. The non-invertible defects of relevance here are constructed by gauging discrete symmetries in a half space. All boundary conditions of two Dirac fermions preserving a $U(1)^2$ symmetry can be constructed by dressing a Dirichlet boundary condition with one of these non-invertible topological lines and sometimes an additional invertible line. These come in two (SPT) classes, one of which (Class $\cV$) are simple conformal boundary conditions while the other (Class $\cA$) are interfaces to the Arf SPT. The latter can be dressed with a Majorana mode to yield a non-simple boundary condition.

A similar correspondence between topological lines and boundary conditions arise for the two-dimensional free compact boson \eqref{eq: compact boson}. At rational values of the radius squared, $R^2=p/q$, a conformal boundary condition can be found which preserves $\cG = \Z_p \times \Z_q \subset U(1)_{\text{shift}} \times U(1)_{\text{winding}}$ \cite{Gaberdiel:2001zq}. Correspondingly, at this value of $R$ the theory is self-dual under gauging $\cG$ and this boundary condition can be constructed by stacking a trivial Dirichlet boundary with the corresponding non-invertible defect. Interestingly, while this $\cG = \Z_p\times \Z_q$ is non-anomalous for all $R$, simple conformal boundary conditions preserving $\cG$ do not exist when $R^2\neq p/q$ \cite{Wei:2025zyd}. This seems to hint that the relation between boundary conditions and non-invertible defects could be more general, which would be a very interesting point to study in the future; e.g.\ by generalising these ideas to interacting theories such as non-linear sigma models \cite{Arias-Tamargo:2025xdd,Arias-Tamargo:2025fhv}.

We have also discussed how symmetric boundary conditions arise in the context of symmetric mass generation in section~\ref{sec:SMG}, showing that a generalisation of the analysis of \cite{Mouland:2025ilu} nicely manifests the difference between the Class $\cV$ and Class $\cA$ boundary conditions of two Dirac fermions. In particular, our construction of boundary conditions preserving symmetries of Class $\cA$ gives only \emph{non-simple} boundaries. Indeed, these symmetries provide yet more examples of anomaly-free symmetries that cannot be preserved by any simple conformal boundary condition.

We finally comment on a few more open questions.

\subsection*{Generalising to $N>2$ fermions}

We have focused on the theory of two Weyl fermions, which is the simplest fermionic theory to exhibit the properties discussed above. Let us consider the generalisation of our analysis to $N>2$ Weyl fermions. 

Interestingly, there are several qualitative differences with respect to the $N=2$ case. Firstly, for $N\ge 8$, this is not the unique CFT with $(c,\bar{c})=(0,N)$ and it is possible that a generic discrete gauging will not result in a self-duality without further fine-tuning. 
Even for $N<8$ there are novelties: we have seen that for $N=2$ there is always a choice of duality isomorphism $D$ such that the corresponding topological line $\cN$ furnishes a Tambara--Yamagami fusion category. All other lines considered can be built by stacking invertible lines on $\cN$. Already for $N=3$ this is not necessarily the case. For example, half-space gauging an anomaly-free $\Z_{39}$ symmetry of three Weyl fermions with charges $(1, 5, 13)$ and using a duality isomorphism $D:\cT/\cG \to \cT$ will lead to a topological self-duality defect. However there is no choice of $D$ such that this defect has TY fusion rules.\footnote{This is guaranteed as there is no choice of $D$ mapping the Wilson line $W$ of $\cT/\cG$ to a generator of $\cG$ (see the discussion in section~\ref{subsec:properties_defects_TYness}).} We hope to be able to report on these matters soon.

\subsection*{Fermion-monopole scattering}

One motivation for this work was the relationship between boundary conditions for two-dimensional fermions and the scattering of light fermions off of heavy monopoles in four-dimensional gauge theory \cite{Callan:1982ah,Rubakov:1982fp,Affleck:1993np,vanBeest:2023dbu,vanBeest:2023mbs}. Let us recall the general story here. We begin with a non-abelian gauge theory that admits 't Hooft--Polyakov monopoles. One is then interested in scattering massless fermions (that do not acquire a mass by the Higgs mechanism) at energies far below the Higgs scale.\footnote{In the context of the Standard Model, what we call here the Higgs scale is really the GUT scale, and we are interested in scattering far below this scale, but far above the electroweak Higgs scale, where quarks are effectively massless.} This is effectively captured by the scattering of massless fermion fields off of 't Hooft lines in the Higgs phase low energy effective theory. If the monopole of interest is spherically symmetric, to first approximation we can solve this scattering mode-by-mode in an angular momentum decomposition. If the four-dimensional fermions were chiral, the scattering problem in the lowest angular momentum channel is precisely a boundary CFT problem of two-dimensional complex fermions with a boundary that must preserve an (anomaly-free) symmetry that is generically chiral, and often non-abelian.

In a number of examples it is precisely the kind of boundary conditions constructed in this paper which solve this scattering problem, where from a two-dimensional perspective local excitations become twist operators at the end of \textit{invertible} lines. This is indeed the case for the charge 1 and 2 spherically-symmetric monopoles in the $SU(5)$ GUT \cite{vanBeest:2023mbs}, for instance.
As a final flourish, one lifts back to four dimensions, and seeks a codimension $1$ topological surface corresponding to the invertible line generated in two dimensions. The idea is that the outgoing state lives on the boundary of such a surface, cut open. Often, the necessary surface in four dimensions is non-invertible, despite the fact it reduces to an invertible line in two dimensions.

A pressing open problem is that there are other perfectly consistent monopoles in perfectly consistent theories---such as the charge 3 monopole in the $SU(5)$ GUT---where we still do not know the solution to the two-dimensional scattering problem of the lowest angular momentum modes. This is the case whenever the left-movers and right-movers carry different $N$-alities under some non-abelian symmetry that the boundary should preserve. Such scattering problems only begin once we have $N\ge 8$ fermions in the two-dimensional effective theory. It would therefore be very interesting to see if a generalisation of the present work to $N>2$ fermions could provide a route towards constructing these (even more) exotic boundaries.

Supposing the correct boundary condition in these cases still attaches a topological line, one is led to ask whether the topological lines in this mysterious case are invertible or not. This is closely related to the classification aspect of the present paper. In detail, we have classified all duality defects of two Weyl fermions that can be produced by gauging an \emph{invertible} symmetry, which is equivalent to finding all boundaries for which the topological lines generated are invertible. We learn that all such defects are $\text{TY}(\Z_k)$ with $k$ odd, up to fusion with an invertible line, and all have a UV description as an abelian rotor. The latter class of defects is known not to generate the required dynamics in those problematic four-dimensional set-ups. Thus if our analysis can be generalised to more than two Weyl fermions, we would have succeeded in showing that fermions scattering off such a monopole must become attached to non-invertible lines in the two-dimensional effective description, which should in turn lift to some non-invertible surface in four dimensions. We hope to report on this direction soon.

\subsection*{Acknowledgements}

We are grateful to Chris Hull for collaboration in the early stages of this work. It's also a pleasure to thank T. Daniel Brennan, Andrea Grigoletto, Antonio Santaniello, Sakura Schäfer-Nameki, Shu-Heng Shao, and David Tong for interesting discussions. GAT is supported by the STFC Consolidated Grant ST/X000575/1. PBS is supported by the ERC-COG grant NP-QFT No.~864583 and the MUR-FARE2020 grant No.~R20E8NR3HX. RM is supported by UK
Engineering and Physical Sciences grant EP/Z000106/1. MVCH was supported by a President's Scholarship from Imperial College London and David Tong's Simons Investigator Award. 

\newpage 
\appendix
\addcontentsline{toc}{section}{Appendices}
\addtocontents{toc}{\protect\setcounter{tocdepth}{0}}

\part*{Appendices}

\section{Details of cyclic group classification}\label{app: proofs}

\subsection{Parameterisation of anomaly-free cyclic groups}

In this appendix we determine all equivalences between the anomaly-free cyclic groups $\cG_{(p,q)}$ defined in section~\ref{sec: classifying groups}.

For the purposes of this appendix let us refer to $\cG_{(p,q)}$ with $pq$ even as symmetries of Class A, and those with $pq$ odd as Class B. 
There are then three scenarios to consider: equivalences within each of Classes A and B, and those between them. We treat each case separately.

\subsubsection*{Equivalences amongst Class A}

Suppose $\cG_{(p,q)}\cong \cG_{(p',q')}$, both in Class A. We have
\begin{align}
  p^2 + q^2 = (p')^2 + (q')^2 = k\,,
\end{align}
and there exists $r$ with $\gcd(r,k)=1$ such that
\begin{align}
  p' 	&= r p \mod k \,,		\nn\\
  q' 	&= r q \mod k \,.
  \label{eq: r rel}
\end{align}
It follows then that
\begin{align}
  pp' + qq' &= Ak	\,,		\nn\\
  pq' - qp' &= Bk \,,
\label{eq: thing}
\end{align}
for some integers $A,B$. But we also have that
\begin{align}
  (A^2 + B^2)k^2=( pp' + qq')^2 + (pq' - qp')^2 = (p^2 + q^2)(p'^2 +q'^2) = k^2\,,
\end{align}
and hence,
\begin{align}
  (pp' + qq',pq'-qp')\in \{(\pm k,0),(0,\pm k)\} \,.
\end{align}
Two of these cases give us $(p',q')=\pm (p,q)$, and indeed it is immediate to see that $\cG_{(p',q')}\cong \cG_{(p,q)}$ in this case. The other two cases are $(p',q')=\pm (q,-p)$. Indeed, we can see that $\cG_{(q,-p)}\cong \cG_{(p,q)}$ by exhibiting the following equality of generators,
\begin{align}
  \Big(e^{-2\pi i p/k},e^{-2\pi i q/k}\Big)^p =  \Big(e^{2\pi i q/k},e^{-2\pi i p/k}\Big)^q\,,
\end{align}
where we recall that $p,q$ are each coprime to $k$.

\subsubsection*{Equivalences amongst Class B}

Suppose $\cG_{(p,q)}\cong \cG_{(p',q')}$, both in Class B. We have
\begin{align}
  p^2 + q^2 = (p')^2 + (q')^2 = k\,,
\end{align}
and there exists $r$ with $\gcd(r,k/2)=1$ such that
\begin{align}
  p' 	&= r p \mod k/2	\,,	\nn\\
  q' 	&= r q \mod k/2 \,.
\end{align}
It follows then that
\begin{align}
  pp' + qq' &= \frac{Ak}{2}	\,,		\nn\\
  pq' - qp' &= \frac{Bk}{2}\,,
\end{align}
for some integers $A,B$. But we also have that
\begin{align}
  (A^2 + B^2)k^2=4( pp' + qq')^2 + 4(pq' - qp')^2 = 4(p^2 + q^2)(p'^2 +q'^2) = 4k^2\,,
\end{align}
and hence,
\begin{align}
  (pp' + qq',pq'-qp')\in \{(\pm k,0),(0,\pm k)\} \,.
\end{align}
Two of these cases give us $(p',q')=\pm (p,q)$, and indeed it is immediate to see that $\cG_{(p',q')}\cong \cG_{(p,q)}$ in this case. The other two cases are $(p',q')=\pm (q,-p)$. Indeed, we can see that $\cG_{(q,-p)}\cong \cG_{(p,q)}$ by exhibiting the following equality of generators,
\begin{align}
  \Big(e^{-4\pi i p/k},e^{-4\pi i q/k}\Big)^{p} =  \Big(e^{4\pi i q/k},e^{-4\pi i p/k}\Big)^q\,,
\end{align}
where we recall that $p,q$ are each coprime to $k/2$.

\subsubsection*{Equivalences between classes}

Finally suppose $\cG_{(p,q)}\cong \cG_{(p',q')}$ where without loss of generality we take $\cG_{(p,q)}$ to belong to Class A and $\cG_{(p',q')}$ to belong to Class B. We have
\begin{align}
  p^2 + q^2 = \frac{(p')^2 + (q')^2}{2} = n \,.
\end{align}
There exists $r$ with $\gcd(r,n)=1$ such that
\begin{align}
  p' 	&= r p \mod n \,,		\nn\\
  q' 	&= r q \mod n \,.
\end{align}
It follows then that
\begin{align}
  pp' + qq' &= An		\,,	\nn\\
  pq' - qp' &= Bn\,,
\end{align}
for some integers $A,B$. But we also have that
\begin{align}
    (A^2+B^2)n^2 = ( pp' + qq')^2 + (pq' - qp')^2 = (p^2 + q^2)(p'^2 +q'^2) = 2n^2\,,
\end{align}
and hence,
\begin{align}
(pp'+qq',pq'-qp') \in \{\pm (n,n),\pm(n,-n)\} \,.
\end{align}
One of these options gives us
\begin{align}
    p' = p-q,\quad q' = p+q\,,
\end{align}
with the other three options related by the known equivalences $(p',q')\sim (-p',-q')\sim (q',-p')$.

Note that we indeed have that $\gcd(p-q,p+q)=1$. To see this, let $\gcd(p-q,p+q)=d$. Then, $d$ divides both $(p+q)+(p-q)=2p$ and $(p+q)-(p-q)=2q$. Since $p,q$ are coprime, this means $d\in \{1,2\}$. But $d$ divides the odd number $(p+q)$, and thus $d=1$.

We can finally check explicitly that for $p,q$ coprime with $pq$ even, we have
\begin{align}
\cG_{p,q}\cong \cG_{p-q,p+q}\,,
\end{align}
by exhibiting the following equality of generators
\begin{align}
  \Big(e^{2\pi i q/k},e^{-2\pi i p/k}\Big)^{p+q} =  \Big(e^{2\pi i (p+q)/k},e^{-2\pi i (p-q)/k}\Big)^q\,,
\end{align}
where $k=p^2 + q^2$. Note that for our conclusion to hold, we need this to be an equality of generating elements. This requires $\gcd(q,k)=1$, which we already know is true. But we also need $\gcd(p+q,k)=1$. This is indeed true, as if a prime $d>1$ divides both $(p+q)$ and $k$ then it divides $ (p+q)^2-k=2pq$. Note then that $d$ cannot divide $p$, for if it did then it would also divide $(p+q)-p=q$, and $p,q$ are coprime. Thus $d|2pq$ implies $d=2$, but $(p+q)$ is odd, and we have a contradiction. Thus, $\gcd(p+q,k)=1$ as required.

\subsection{Subgroups of $\cG_{(p,q)}$}

We prove here that any subgroup of $\cG_{(p,q)}$ as defined in section~\ref{sec: classifying groups} is equivalent to $\cG_{(p',q')}$ for some coprime pair $(p',q')$. Due to the equivalences \eqref{eq: A B equivalence} it suffices to consider $pq$ even. Suppose $N$ divides $k=p^2 + q^2$. We want to show that the $\Z_N$ subgroup of $\cG_{(p,q)}\cong \Z_k$ is isomorphic to $\cG_{(p',q')}$ for some coprime pair $(p',q')$ where again we can assume $p'q'$ even.

We can write
\begin{align}
    k = NM \,.
\end{align}
The key step is to show that there exist integers $a,b,c,d$ satisfying all of the following,
\begin{align}
    N &= a^2 + b^2,\qquad \gcd(a,b) = 1 \,,\nn\\
    M &= c^2 + d^2,\qquad \gcd(c,d)=1   \,,\nn\\
    p &= ac-bd     \,, \nn\\
    q &= ad + bc \,.
\label{eq: subgroup claim}
\end{align}
Note that $N$ and $M$ are odd, and thus $ab$ and $cd$ are even.

The first step in demonstrating \eqref{eq: subgroup claim} is to show that any prime divisor $d$ of $k$ satisfies $d=1$ mod $4$. So let $d$ be a prime divisor of $k$. Since $k$ is odd, so is $d$. Furthermore $\gcd(p,q)=1$ implies that $\gcd(p,d)=\gcd(q,d)=1$.\footnote{This is easily seen by contradiction. Assume $g\equiv \gcd(p,d)>1$ then $g|p \implies g|p^2$, while $g|d$ and $d|k$ imply that $g|q^2$, so $g$ divides both $p^2$ and $q^2$, which contradicts $\gcd(p,q)=1$.} We therefore have
\begin{equation}
    k = 0\mod d \;\implies\; p^2 = -q^2 \mod d \;\implies\; (q^{-1}p)^2 = -1 \mod d \,,
\end{equation}
where $q^{-1}$ is the multiplicative inverse of $q$ modulo $d$, which is well-defined as $\gcd(q,d)=1$. Therefore, $-1$ is a quadratic residue modulo $d$. This implies that $d=1\mod 4$.\footnote{For an odd prime $d$, if $\exists x$ such that $x^2 = -1 \mod d$ then $(x^2)^{(d-1)/2} = x^{d-1} = (-1)^{(d-1)/2} \mod d$. Fermat's little theorem implies that $x^{d-1} = 1 \mod d$ since $x\neq 0\mod d$. Therefore $(-1)^{(d-1)/2}=1\mod d$, implying that $(d-1)/2$ is even, so $d=1\mod 4$.}

We can therefore perform the prime factorisation
\begin{align}
    k = p^2 + q^2 = u_1\dots u_n v_1\dots v_m\,,\qquad N=u_1\dots u_n,\,\, M = v_1\dots v_m\,,
\label{eq: k prime factorisation}
\end{align}
for (not necessarily distinct) primes $u_\alpha,v_\mu=1$ mod $4$, where here $\alpha=1,\dots,n$ and $\mu=1,\dots,m$.

To proceed, it is useful to think in terms of the Gaussian integers\footnote{A useful reference text that includes proofs of a few facts we use below is \cite{Conrad2013THEGI}.} $\Z[i]=\Z+i\Z$. This is a commutative ring whose units are $u=1,-1,i,-i$. We define the norm $\Upsilon(a+ib) = (a+ib)(a-ib) = a^2 + b^2$, which is multiplicative: $\Upsilon(\sigma\tau) = \Upsilon(\sigma)\Upsilon(\tau)$ for $\sigma,\tau\in \Z[i]$.

Consider then the prime factorisation of $(p+iq)$ over $\Z[i]$. The primes $\sigma$ of $\Z[i]$ are of three types:
\begin{enumerate}[label=(\roman*)]
    \item $\sigma = u(1+i)$ for unit $u$. These have $\Upsilon(\sigma)=2$.
    \item $\sigma$ such that $\Upsilon(\sigma)=p$ for $p\in \mathbb{N}$ prime with $p=1$ mod $4$.
    \item $\sigma=up$ for unit $u$ and $p\in \mathbb{N}$ prime with $p=3$ mod $4$. These have $\Upsilon(\sigma)=p^2$.
\end{enumerate}
Note then that \eqref{eq: k prime factorisation} provides a prime factorisation over $\Z$ of the norm $\Upsilon(p+iq) = p^2 + q^2$, which implies that the prime factorisation of $(p+iq)$ over $\Z[i]$ contains only primes of type (ii). Indeed, we can write
\begin{align}
(p+iq) = \sigma_1\dots \sigma_n \tau_1\dots \tau_m\,,
\end{align}
with $\sigma_\alpha,\tau_\mu\in \Z[i]$ prime obeying
\begin{align}
\Upsilon(\sigma_\alpha) = u_\alpha\,,\qquad \Upsilon(\tau_\mu) = v_\mu \,.
\end{align}
So finally we can define $a,b,c,d\in \Z$ by
\begin{align}
(a+ib) = \sigma_1\dots \sigma_n,\qquad (c+id)= \tau_1\dots \tau_m\,,
\end{align}
so that
\begin{align}
    (p+iq) = (a+ib)(c+id) \,.
\end{align}
Any common factor of $a,b$ is also a common factor of $p,q$, and similarly for $c,d$, so that $\gcd(a,b)=\gcd(c,d)=1$. It is then straightforward to show that $a,b,c,d$ also satisfy all other properties listed in \eqref{eq: subgroup claim}.

Then given $a,b,c,d$ satisfying \eqref{eq: subgroup claim}, we have
\begin{align}
\Big( e^{2\pi i q/N}, e^{-2\pi i p/N} \Big)^b = \Big( e^{2\pi i b/N}, e^{-2\pi i a/N} \Big)^q \,.
\end{align}
Since both $b$ and $q$ are coprime to $N$, this establishes that the subgroup $\Z_N\subset \cG_{(p,q)}$ is isomorphic to $\cG_{(a,b)}$, and thus we have proven the claim with $(p',q')=(a,b)$. Note that as shown above, $p'q'=ab$ is even.

\section{Self-duality for other spin structures}\label{app: RR_duality}

In section~\ref{subsec: self-duality} we demonstrated self-duality of $\cT$ under gauging $\cG=\cG_{(p,q)}$ in the NS-NS sector. We provide here the details of the generalisation of this computation to other spin structures, the results of which are summarised in \eqref{eq:general duality}. We leave the choice of $C$ used in the map \eqref{eq:action_C}, generic. The possible choices considered in the main text are given in \eqref{eq:four_Cs}.

\subsection{The NS-NS sector}

We saw that acting with the map \eqref{eq: reference iso}, we obtain in the NS-NS sector
\begin{equation}
	\hat{Z}^{(0,0)}(\tau,\mu_1,\mu_2) = Z^{(0,0)}(\tau,\mu_2,\mu_1) \,.
\end{equation}
The generalisation of this expression for the three other choices of $C$ appearing in \eqref{eq:four_Cs} is straightforward. By acting with \eqref{eq:action_C}, we find
\begin{align}
  \hat{Z}^{(0,0)}(\tau,\vec{\mu}) = Z^{(0,0)}(\tau,K \vec{\mu})\,,
  \label{eq: NS-NS starting}
\end{align}
where in this appendix it will be useful to use vector notation for chemical potentials $\vec{\mu} = (\mu_1,\mu_2)$. The matrix $K$ is given in \eqref{eq:four K's} corresponding to the four choices of $C$ in \eqref{eq:four_Cs}.

\subsection{The NS-R sector}

We then have in the NS-R sector
\begin{align}
  \hat{Z}^{(0,1)}(\tau,\vec{\mu})		&= \hat{Z}^{(0,0)}(\tau,\vec{\mu} + \vec{\nu}) 		\nn\\
  										&= Z^{(0,0)}(\tau, K(\vec{\mu} + \vec{\nu}))			\nn\\
  										&= Z^{(0,1)}(\tau, K \vec{\mu} + (K - 1) \vec{\nu})\,,
\end{align}
where here $\vec{\nu} =  (1,2q-1)\pi i$ and we have used \eqref{eq: sector shift T/G}, \eqref{eq: sector shift T} and \eqref{eq: NS-NS starting}. 

To proceed, we need to take into account the periodicity conditions obeyed by $Z^{(u,v)}(\tau,\vec{\mu})$, by virtue of the global form of the symmetry group. Defining 
\begin{align}
  \vec{\sigma}_1 = 2\pi i (1,0),\qquad \vec{\sigma}_2 = 2\pi i(0,1),\qquad \vec{\sigma}_3 = \frac{2\pi i }{k}(q,-p)\,,
\end{align}
we have in all
\begin{align}
  Z^{(u,v)}(\tau,\vec{\mu} + a_1 \vec{\sigma}_1  + a_2 \vec{\sigma}_2  + a_3 \vec{\sigma}_3) = (-1)^{u(a_1+a_2+a_3)}Z^{(u,v)}(\tau,\vec{\mu}) \,,
\label{eq: chem pot shifts}
\end{align}
for $a_1,a_2,a_3\in \Z$.
We then find
\begin{align}
  (K-1)\vec{\nu} = \{&(q-1)\vec{\sigma}_1 + (1-q)\vec{\sigma}_2,-q\vec{\sigma}_1-q\vec{\sigma}_2\nn\\
  &(1-2q)\vec{\sigma}_2 + (p-q-2pq)\vec{\sigma}_3,-\vec{\sigma}_1 - (p-q-2pq)\vec{\sigma}_3\}\,,
\label{eq: K shift values}
\end{align}
in each of the four cases. Thus, using \eqref{eq: chem pot shifts} we find in every case
\begin{align}
  \hat{Z}^{(0,1)}(\tau,\vec{\mu})		&= Z^{(0,1)}(\tau, K \vec{\mu} + (K - 1) \vec{\nu}) \nn\\
  										&= Z^{(0,1)}(\tau, K \vec{\mu} ) \,,
\end{align}
as claimed.

\subsection{The R-NS sector}

The partition function of $\cT/\cG$ in the R-NS sector is given by
 \begin{align}
    &\hat{Z}_{(\alpha,\beta)}^{(1,0)}(\tau,\vec{\mu}) = \frac{1}{\overline{\eta(\tau)}^2} \sum_{n_1,n_2\in \Z-\frac{1}{2}} \bar{\zeta}^{(n_1+\alpha q/k)^2/2 +(n_2-\alpha p/k)^2/2} z_1^{pn_1+qn_2}z_2^{qn_1 - pn_2 + \alpha }		\nn\\
    &\hspace{35mm} \times \exp\left(\frac{2\pi i \beta}{k} \big(qn_1 - pn_2+t_* p\alpha\big)- \pi i \beta\right)		\,.
\label{eq: twisted PF R-NS}
\end{align}
The sum over $\beta$ projects onto states $(n_1,n_2,\alpha)$ satisfying
\begin{align}
  qn_1 - pn_2 + t_* p \alpha -\frac{k}{2} = 0\mod k\,,
\end{align}
where we emphasise that the left-hand-side of this equation is indeed an integer, since $n_1,n_2\in \Z+\frac{1}{2}$ and one of $p,q$ is even. This constraint is then equivalent to 
\begin{align}
 \alpha - 2(pn_2-qn_1) = 0 \quad \mod k \,.
\end{align}
We can parameterise solutions to this equation with $m_1,m_2\in \Z+\frac{1}{2}$ defined by
\begin{align}
m_1(\alpha,n_1,n_2) &= n_1-
  \frac{q}{k}\big(  \alpha - 2(pn_2-q n_1) \big) \,,         \nn\\
m_2(\alpha,n_1,n_2) &= n_2+
  \frac{p}{k}\big( \alpha - 2(pn_2-q n_1)  \big) \,.
\end{align}
This can be written equivalently in the compact form
\begin{align}
  \begin{pmatrix}
  	m_1(\alpha,n_1,n_2) \\ m_2(\alpha,n_1,n_2)
  \end{pmatrix} = \frac{1}{k} \begin{pmatrix}
  	a & b \\ b & -a
  \end{pmatrix}\begin{pmatrix}
  	n_1 + \frac{\alpha q}{k} \\ n_2 - \frac{\alpha p}{k}
  \end{pmatrix}\,,
\end{align}
with $a,b$ defined in \eqref{eq: Euclid}. This is indeed a one-to-one parameterisation, as evidenced by the inverse map
\begin{align}
\alpha(m_1,m_2)   &= 2(pm_2-qm_1)\quad (\text{mod }k)\,, \,  \qquad \alpha(m_1,m_2)\in \{0,1,\dots,k-1\}\,,      \nn\\
n_1 (m_1,m_2)     &= m_1 - \frac{q}{k}\big( \alpha(m_1,m_2) - 2(pm_2-q m_1) \big) \,,  \nn\\
n_2 (m_1,m_2)     &= m_2 + \frac{p}{k}\big( \alpha(m_1,m_2) - 2(pm_2-q m_1) \big) \,. 
\end{align}
As for the NS-NS sector in \eqref{eq: reference iso}, we can generalise this map by the replacement
\begin{align}
  \begin{pmatrix}
  	m_1 \\ m_2
  \end{pmatrix} \to C \begin{pmatrix}
  	m_1 \\ m_2
  \end{pmatrix} \,,
\end{align}
with $C$ one of the four matrices \eqref{eq:four_Cs}. Importantly, after this replacement $m_1,m_2$ still belong to $\Z+\frac{1}{2}$. 

Plugging $n_1(m_1,m_2),n_2(m_1,m_2)$ into \eqref{eq: twisted PF R-NS} we land straightforwardly on the claimed result,
\begin{align}
  \hat{Z}^{(1,0)}(\tau,\vec{\mu})	&= Z^{(1,0)}(\tau, K \vec{\mu} ) \,.
\label{eq: R-NS duality}
\end{align}

\subsection{The R-R sector}

Starting with \eqref{eq: R-NS duality} we have
\begin{align}
  \hat{Z}^{(1,1)}(\tau,\vec{\mu}) 		&= \hat{Z}^{(1,0)}(\tau,\vec{\mu} + \vec{\nu})		\nn\\
  										&= Z^{(1,0)}(\tau, K(\vec{\mu} + \vec{\nu}))			\nn\\
  										&= Z^{(1,1)}(\tau, K \vec{\mu} + (K - 1) \vec{\nu})\,,
\end{align}
where we have again used \eqref{eq: sector shift T/G} and \eqref{eq: sector shift T}. We can now plug in the values \eqref{eq: K shift values} for $(K-1)\vec{\nu}$ in each of the four cases and apply \eqref{eq: chem pot shifts} to find
\begin{align}
  \hat{Z}^{(1,1)}(\tau,\vec{\mu}) 		&= Z^{(1,1)}(\tau, K \vec{\mu}) \,,
\end{align}
which takes the desired form \eqref{eq:general duality}. Indeed, this simplifies to \eqref{eq: reference_duality} for the first choice of $C$ in \eqref{eq:four_Cs}.

The $O(4)$ matrices corresponding to each choice of $C$ in \eqref{eq:four_Cs} are
\begin{align}
  \left\{
  \begin{pmatrix}
0 & \sigma_3 \\
\mathds{1} & 0
\end{pmatrix}\,,
  \begin{pmatrix}
0 & \mathds{1} \\
\sigma_3 & 0
\end{pmatrix}\,,
  \begin{pmatrix}
\mathds{1} & 0 \\
0 & \sigma_3
\end{pmatrix}\,,
  \begin{pmatrix}
\sigma_3 & 0 \\
0 & \mathds{1}
\end{pmatrix} 
\right\}\,,
\end{align}
where $\sigma_3$ here is the third Pauli matrix. As all of these have determinant $-1$, it is clear that they are elements of $O(4)$ and not $SO(4)$. That is, from the discussion at the end of section~\ref{subsec: self-duality}, these do not correspond to symmetries of $\cT$ but instead give isomorphisms from $\cT$ to $\cT\otimes\text{Arf}$.

\section{Tambara--Yamagami bicharacter}\label{app: bicharacter}

The goal of this appendix is to elaborate on the structure of the two Tambara--Yamagami lines (defined via the first two matrices in \eqref{eq:four_Cs}). In section~\ref{sec:fusion_rules} we have discussed the difference between the fusion rules of these lines and those of the lines corresponding to the final two matrices in \eqref{eq:four_Cs}. The full categorical data includes more information, namely the $F$-symbols that control the associativity properties of topological defects. In the case of a TY category, the $F$-symbols are completely determined by three pieces of data \cite{Tambara:1998vmj, Thorngren:2021yso}: the order of the cyclic group $\cG \cong \Z_k$ generated by the $\eta$ line, the object known as the \emph{symmetric bicharacter} (which we discuss in this appendix), and a sign $\epsilon=\pm 1$ known as the Frobenius--Schur indicator. We will not discuss the latter here.

As stated in the main text, the defect $\cN$ obeys the fusion rules of a TY fusion category precisely when the isomorphism $D$ between $\cT/\cG$ and $\cT$ maps the dual symmetry $\widehat{\cG} \cong \Z_k$ of $\cT/\cG$ onto the symmetry $\cG$ of $\cT$. Let us assume this, so that $C$ takes one of the first two values in \eqref{eq:four_Cs}. The way in which the elements of $\widehat{\cG}$ are identified with those of $\cG$ is encoded in the integer $r$ with $\gcd(r,k)=1$ found in \eqref{eq: W eta new}.
It is this integer which fixes the \textit{symmetric bicharacter} $\chi$ of the TY fusion category. In general, $\chi$ specifies how background gauge fields coupled to $\cG$ in $\cT$ are related to those coupled to $\widehat{\cG}$; that is, schematically we have
\begin{align}
    \sum_A Z[A] e^{i\chi[A,A']} = Z[A'] \,.
\label{eq: bichar general?}
\end{align}
Let us be more explicit. In our case, the bicharacter can be written $e^{i\chi(A,A')} = \omega^{\int A \smile A'}$ where $\omega$ is a primitive $k^\text{th}$ root of unity. In terms of $r$, it is given by
\begin{align}
  \omega=e^{-2\pi i r/k} \,.
\end{align}
The isomorphism $D:\cT/\cG \to \cT$ holds in the presence of an arbitrary background $\widehat{\cG}$ gauge field, mapping it to a particular $\cG$ gauge field. Then $\omega$ can be \textit{defined} as the unique primitive $k^\text{th}$ root of unity such that 
\begin{align}
 \frac{1}{k}\sum_{\alpha,\beta=0}^{k-1} \hat{Z}^{(u,v)}_{(\alpha,\beta)}\omega^{(\alpha \beta' - \alpha' \beta)} = \hat{Z}^{(u,v)}_{(\alpha',\beta')}\,,
\label{eq: bichar}
\end{align}
where, as in the text, $\hat{Z}^{(u,v)}_{(\alpha,\beta)}$ is the twisted torus partition function \eqref{eq: twisted PF} (with $t=t_*$). On the right-hand side we have coupled $\cT$ to a background gauge field $A'$ for $\cG$ which is specified by holonomies $(\alpha',\beta')$. Meanwhile, on the left-hand-side $rA'$ is the background field for the dual symmetry $\widehat{\cG}$. This is achieved through a coupling to the dynamical gauge field $A$, specified by holonomies $(\alpha,\beta)$, given by
\begin{align}
  \omega^{\int A \smile A'} = \omega^{(\alpha\beta'-\alpha'\beta)} \,.
\end{align}
In section~\ref{subsec:properties_defects_TYness}, we found that \eqref{eq: bichar} is satisfied with $r=\pm (2q)^{-1} p$, with sign depending on which $D$ we've picked, for the special case $(\alpha',\beta') = (0,\pm 2q)$ and $(u,v)=(0,0)$. For completeness, we now verify \eqref{eq: bichar} for general $A'$, still for $(u,v)=(0,0)$. 

Let's specialise to the isomorphism $D$ corresponding to the first $C$ matrix in \eqref{eq:four_Cs}. Picking the opposite sign for $C$ follows similarly. On the left-hand side of \eqref{eq: bichar} we then have
\begin{align}
\begin{split}\label{eq: NS-NS bichar}
\frac{1}{k} \sum_{\alpha,\beta=0}^{k-1} \frac{1}{\overline{\eta(\tau)}^2} &\sum_{n_1,n_2\in \Z} \bar{\zeta}^{(n_1+\alpha q/k)^2/2 +(n_2-\alpha p/k)^2/2} \\
& \times \;\exp\left(\frac{2\pi i \beta}{k} \left(qn_1 - pn_2+ t_*p\alpha +r \alpha'\right) - \frac{2\pi ir\alpha \beta'}{k} \right) \,.
\end{split}
\end{align} 
The sum over $\beta$ projects onto states with
\begin{align}
qn_1 - pn_2+ t_*p\alpha +r \alpha' = 0 \mod k \,,
\end{align}
or, equivalently, those with
\begin{align}
 \alpha + 2r \alpha' - 2( pn_2 - qn_1) = 0 \mod k \,.
\label{eq: constraint alpha prime}
\end{align}
We can parameterise solutions $(n_1,n_2,\alpha)$ of this equation by $(m_1,m_2)\in \Z^2$ with
\begin{align}
\alpha(m_1,m_2)   &= -2r \alpha' + 2(pm_1+qm_2)\quad (\text{mod }k)\,, \,  \qquad \alpha(m_1,m_2)\in \{0,1,\dots,k-1\}\,,      \nn\\
n_1 (m_1,m_2)     &= -m_2 - \frac{q}{k}\big( \alpha(m_1,m_2)+2r\alpha' - 2(pm_1+qm_2) \big) \,,  \nn\\
n_2 (m_1,m_2)     &= m_1 + \frac{p}{k}\big( \alpha(m_1,m_2) +2r\alpha' - 2(pm_1+qm_2) \big) \,. 
\end{align}
This is a one-to-one parameterisation of solutions to \eqref{eq: constraint alpha prime}, as evidenced by the inverse map
\begin{align}
-m_2(\alpha,n_1,n_2) &= n_1-
  \frac{q}{k}\big(  \alpha +2r\alpha'- 2(pm_1+qm_2) \big) \,,         \nn\\
m_1(\alpha,n_1,n_2) &= n_2+
  \frac{p}{k}\big( \alpha+2r\alpha' - 2(pm_1+qm_2)  \big) \,.
\end{align}
The dimension of the operator $(m_1,m_2)$ is then given by
\begin{align}
  \frac{1}{2}\left(n_1 + \frac{\alpha q}{k}\right)^2 + \frac{1}{2}\left(n_2 - \frac{\alpha p}{k}\right)^2 &= \frac{1}{2}\left(m_1-\frac{2pr\alpha'}{k}\right)^2 + \frac{1}{2}\left(m_2-\frac{2qr\alpha'}{k}\right)^2 \nn\\
  &= \frac{1}{2}\left(m_1' + \frac{\alpha' q}{k}\right)^2 + \frac{1}{2}\left(m_2' - \frac{\alpha' p}{k}\right)^2\,,
\end{align} 
where we've defined shifted mode numbers
\begin{align}\label{eq:shifted modes_bich}
  m_1' &= m_1 + \frac{(-2pr-q)\alpha'}{k} \,,		\nn\\
  m_2' &= m_2'+\frac{(-2qr+p)\alpha'}{k} \,.
\end{align}
We require $m_1',m_2'\in \Z$ for all $\alpha'$, which is the case precisely if
\begin{align}
  r = (2q)^{-1}p \mod k \,.
  \label{eq: r result}
\end{align}
Lastly, we need to evaluate the final term appearing in the exponential in \eqref{eq: NS-NS bichar}. We find
\begin{align}
 -r\alpha(m_1,m_2) &=  qm_1-pm_2 + 2r^2 \alpha' \mod k		\nn\\
  &= qm_1-pm_2 + t_*p\alpha' \mod k			\nn\\
  &= qm_1'-pm_2' + t_*p\alpha' \mod k	\,,
\end{align}
using $t_*p = 2^{-1}$ mod $k$. Hence
\begin{align}
\exp \left( -\frac{2\pi ir\alpha \beta'}{k} \right) &=\exp \left( \frac{2\pi i\beta'}{k}\Big(  qm_1' - pm_2' +t_* p\alpha' \Big)\right) \,.
\end{align}
Putting the pieces together, with $r$ given by \eqref{eq: r result}, we find
\begin{align}
  &\frac{1}{k}\sum_{\alpha,\beta=0}^{k-1} \hat{Z}_{(\alpha,\beta)}\omega^{(\alpha \beta' - \alpha' \beta)}	\nn\\
  &\qquad   = \frac{1}{\overline{\eta(\tau)}^2} \sum_{m_1',m_2'\in \Z} \bar{\zeta}^{(m_1'+\alpha' q/k)^2/2 +(m_2'-\alpha' p/k)^2/2}\exp\left(\frac{2\pi i \beta'}{k} \left(qm_1' - pm_2'+ t_*p\alpha'\right)\right)	\nn\\
  &\qquad =\hat{Z}_{(\alpha',\beta')} \,,
\end{align}
as claimed.

\bibliographystyle{JHEP}
\bibliography{refs}

\end{document}